\documentclass{article}

\usepackage[preprint]{neurips_2026}
\usepackage{hyperref}       %
\usepackage{url}            %
\usepackage{booktabs}       %
\usepackage{amsfonts}       %
\usepackage{nicefrac}       %
\usepackage{microtype}      %
\usepackage{xcolor}         %

\usepackage{subcaption}
\usepackage{tabularx}
\usepackage{graphicx} %

\usepackage{multirow}
\usepackage{enumitem}

\usepackage{amsmath}
\usepackage{amssymb}
\usepackage{mathtools}
\usepackage{amsthm}
\usepackage{placeins}
\usepackage{algorithm}
\usepackage{algorithmic}
\usepackage{bbm}

\usepackage{wrapfig}

\newcommand{\eat}[1]{}

\usepackage[capitalize,noabbrev]{cleveref}

\theoremstyle{plain}

\theoremstyle{definition}

\theoremstyle{remark}

\newcommand{\yrcite}[1]{\citeyearpar{#1}}

\title{Room for Error: Large-Scale Simulation of Over-the-Air Acoustic Attacks}

\author{%
  Andrew C. Cullen \\
  University of Melbourne \\
  \texttt{andrew.cullen@unimelb.edu.au} \\
  \And
  Neil G. Marchant \\
  University of Melbourne \\
  \And
  Jiani Xie \\
  University of Melbourne 
  \AND 
   Paul Montague \\
   DST Group, Adelaide \\
   \And
      Sean Lamont \\
   DST Group, Adelaide \\
   \And 
   Maxwell Standen \\
   DST Group, Adelaide \\
   \And
  Benjamin I. P. Rubinstein \\
  University of Melbourne \\
}

\begin{document}

\maketitle

\begin{abstract}
While voice control is rapidly becoming a ubiquitous vector of human-AI communication, the risks facing these systems remain poorly understood. This is, in part, a product of the difficulties in scaling strictly digital adversarial workflows to the physical world. These scale barriers have led the community to abstract away key acoustic factors relating to detectability and the influence of geometry on acoustics. These methodological and metrological shortcomings undermine our understanding of risk. We illuminate these issues through real-world testing, conceptual discussions, and a novel, high-throughput reality simulation framework. By testing over 8 million adversarial evaluations, we demonstrate that acoustic awareness yields relative Word Error Rate increases of up to $94.5\%$ under Whisper and wav2vec. We employ this framework to explore a formalize and operationalize a \textit{Dual-Form Signal to Noise Ratio} to decouple source stealth from victim attack efficacy, resolving a crucial limitation in current works. This lays the groundwork for repeatable, verifiable research that embraces, rather than abstracts, the acoustic environment.

\end{abstract}

\section{Introduction}

Historically, the AI risk community has primarily targeted computer vision systems, based upon the observation that these systems are susceptible to adversarial perturbations---carefully crafted perturbations that, while imperceptible to humans, elicit anomalous behavior in models~\citep{goodfellow2014explaining, carlini2017towards}. However, the landscape of AI systems has fundamentally changed since these early works, with text and voice user interfaces emerging as the central form of AI interaction. This in turn shifts the risk landscape towards these models, especially as they are increasingly exposed to sensitive or private scenarios, due to their integration in everything from smart devices~\citep{zhang2024laseradv} to AI scribes~\citep{draper2025clinical}. As such, it is crucial that we understand the risks associated with the Automatic Speech Recognition (ASR) models that underpin these interactions in order to inhibit potential harms.%

Carlini \& Wagner \yrcite{carlini2018audio} were the first to establish that Deep Neural Networks for ASR are susceptible to harms induced by inaudible adversarial perturbations, causing erroneous transcriptions or malicious command execution. However, much of the subsequent literature focuses upon signals that exist strictly in the digital domain---where the sample, perturbation, attacker, and model all exist as digital audio signals, and where each party holds perfect information about the domain. 
This assumes a threat model where recorded audio can be intercepted and modified prior to model access, implying a level of device compromise that would obviate the need for adversarial examples.%

More generally, even when ASR systems ultimately consume digital audio, these signals are typically the end product of a physical process. Speech is emitted into an acoustic environment and only later digitized for model consumption. This physical channel introduces reverberation, attenuation, and interference that fundamentally shape the signal observed by the ASR system. Abstracting away this process amounts to assuming unrealistic adversarial access at the level of the system's digital input, rather than influence mediated by the physical world.

This is not to say that real-world (as opposed to strictly digital) attacks have not taken place; indeed, many such works arise precisely in response to the physical considerations above. Over-the-Air (OTA) attacks have been proposed, where the perturbation must survive playback through speakers, propagation through air, and recording by microphones~\citep{yuan2018commandersong, li2020advpulse, qin2019imperceptible, szurley2019perceptual, olivier2022there}. However these explorations do not consider the acoustic environment as anything more than just a lossy map drawn from a single physical configuration. In ignoring the influence of the broader physical domain, these approaches end up being both impossible to replicate and highly predicated upon the specific physical configuration they were tested within, limiting both scientific validity and our broader understanding of OTA risks.

On the one hand, digital only attacks abstract away most of the parts that make acoustic environments interesting, including attenuation, reflections, and the presence of distinct, physically located attackers, victims, and audio sources. On the other, OTA attacks are difficult to implement and are predicated upon singular acoustic environments---essentially abstracting away the very same acoustic information. This is inherently problematic, as the physical acoustic space is a harsh environment for adversarial perturbations. It acts as a complex, frequency-selective filter characterized by reverberation, attenuation, and background noise. While research has attempted to somewhat resolve these issues~\citep{athalye2018synthesizing, chen2023advreverb, zhang2024advsv}, we have identified three critical flaws in the current evaluation methodologies for OTA attacks:
\begin{enumerate}[leftmargin=*, noitemsep, topsep=0pt]
    \item \textbf{The SNR Ambiguity:} Literature consistently reports attack stealth using a single Signal-to-Noise Ratio (SNR) scalar (e.g., ``$20$~dB''). In a physical room, this is an ill-defined unless the measurement location is specified, and opens additional questions. Should this ratio exist at the victim's microphone (where the attack must overpower the clean signal to succeed) or at the attacker's speaker (where the attack must be quiet to remain undetected by the original speaker)? We argue that it is crucial that the community begins to standardize acoustic-awareness in metrics.    
    
    \item \textbf{The White-Box Geometry Fallacy:} Many advanced attacks assume the attacker can either ignore, or precisely model, the room's acoustics properties to craft their perturbation. In practice, these properties are key for attack performance, yet an attacker will rarely have the opportunity to perform a sufficient sine-sweep of a victim's acoustic environment to accurately measure them, let alone to precisely model microphones at the source and/or target. The degradation of attack success as this geometric knowledge fades is poorly understood.
    
    \item \textbf{Statistical Insignificance:} Due to the high experimental cost of physical evaluation, experiments often consider a single acoustic environment, failing to capture the span of real-world environments.
\end{enumerate}

\textbf{Our Contributions.} To address these gaps, our work makes the following contributions:
\begin{enumerate}[leftmargin=*, noitemsep, topsep=0pt]
    \item We formalize the \textbf{Knowledge Gradient}, a continuous spectrum of attacker information ranging from naive digital attacks to perfect white-box omniscience, exposing the ``Information Cost" of room uncertainty. 
    \item We introduce \textbf{Dual-Form SNR Metrics} to explicitly decouple source stealth from target interference, revealing the energetic cost of projection often hidden in single-scalar reporting, formalising a preliminary approach initially articulated by Abdullah et al. \yrcite{abdullah2021sok}.
    \item We provide the foundations for \textbf{high-throughput simulations of acoustically-aligned environments}, to support modeling of the impacts of simulated, known or unknown RIRs as part of a broader acoustic attack infrastructure. 
\end{enumerate}
These results are supplemented by a novel formalism of acoustic adversarial attacks, that allows us to articulate the opportunities for significant new, acoustically-aligned research in the Adversarial Machine Learning (AML) space.

\section{Acoustic Attacks}\label{sec:acoustic_attacks}

Formally, an acoustic attack against an ASR system can be defined as follows. Let $\mathcal{X} \subseteq \mathbb{R}^T$ denote a fixed length $T$ stored representation of an acoustic waveform and $\mathcal{Y}$ the space of token sequences. An ASR model is then the mapping $
M : \mathcal{X} \to \mathcal{Y},$ which assigns to an input waveform $x \in \mathcal{X}$ a transcription $y = M(x)$. An adversarial attack seeks a perturbation $\delta \in \mathcal{X}$ such that the perturbed transcription $y' = M(x + \delta)$ diverges from $y$, typically measured via the Word Error Rate $\operatorname{WER}(y, y')$ through either targeted or untargeted mechanisms.

We can generalize this scenario to consider the \emph{information available to the attacker} through the information operator $H : \mathcal{X} \to \mathcal{Z},$ where $\mathcal{Z}$ denotes the attacker's observation space and $H(x)$ denotes the observation of the waveform available to the attacker. This formulation allows us to taxonomically classify attacks into three key classes: the case where $H(x) = x$, denoting perfect information; $H(x) = \emptyset$, where there is no sample specific information, leading to universal attacks; or $H(x; t) = (x_1, \ldots, x_t)$; a partial information case where the attacker observes a prefix of the waveform up to time $t$. The attack is also classified in terms of its strength, limiting $\delta$ to within some measure---typically based upon the Signal-to-Noise Ratio (SNR) 
\begin{equation}\label{eqn:SNR}
\mathrm{SNR}(\delta, x) = 10 \log_{10}\!\bigl(\|x\|_p^p \,/\, \|\delta\|_p^p\bigr) \;\ge\; \mathrm{SNR}_{\text{tgt}},
\end{equation}
for some $\ell_p$ norm (typically $p \in \{2, \infty\}$) or perceptual measures~\citep{qin2019imperceptible, schonherr2019psychoacoustic_hiding, szurley2019perceptual, sun2024commanderuap}. While some of these perceptual measures have been employed in prior works, they were designed to assess speech quality and intelligibility under benign channel distortions, rather than to quantify adversarial impact or detectability.
Disconcertingly, all of these measures abstract away the physical nature of acoustic environments, in that they ignore the influence of relativity in the context of the physical positions of the source $p_{s}$, the target/victim $p_{v}$, and the attacker $p_{a}$.

\paragraph{Relativity of Observations} These limitations speak to the background of those performing adversarial machine learning---where expertise is concentrated upon the mechanics of machine learning---and the difficulties associated with real-world environmental testing. However, while drawing insights from computer vision style adversarial attacks helped provide the foundations of audio AML, it is also obfuscates the need for solving crucial, domain specific research tasks.

A key example of this relates to the geometric nature of sound, and its existence over an acoustic space. While this has been noted in multiple papers including  \cite{abdullah2021sok}, the evaluation of audio attacks has almost exclusively ignored the difference between the sources of acoustic signals (be they the original audio source or the attacker), and the nature of perception across different parts of the acoustic space, and how this attenuates the information available to the victim. This leads to assuming that $H$ transfers a clean version of $x$.

Within this work, we explicitly consider SNR as a property at both the victim/target, which is correlated with attack performance, and at the source, which correlates with detectability. While it is true that target detectability may also be possible (especially using automated systems), we focus on the implicit threat model behind ASR research, in which the highest likelihood of detectability involves the original speaker observing the modified signal. 

As such, we instead consider $H$ to be a lossy physical transfer function $H_{p_{i} \to p_{j}}$ from a source location $p_{i}$ to a target location $p_{j}$, that depends upon the environment $\omega$ and measurement infrastructure $\psi$. This reflects the fundamental physical reality, that there exists a \textbf{spatial mismatch} between the source transmission and both the attacker and victim observations of the signal. This in turn makes it clear that there is not one single constraint on the SNR, but potentially multiple, as the perturbed signal risks detection at both the source and victim. We capture this by introducing the \textbf{Dual-Form SNR}, which considers the SNR at the source ($\text{SNR}_{s}$) and victim ($\text{SNR}_{v}$), where we replace Equation~\ref{eqn:SNR} with 
\begin{equation}
    \text{SNR}_{i} = 10 \log_{10} \left( \frac{P_{H_{p_s \to p_i}(x)}}{P_{H_{p_a \to p_i}(\delta)}} \right) \qquad i \in \{s, v\}
\end{equation}
where $P$ represents the mean squared power. 

If the SNR is treated as a proxy for stealth, then evaluating stealth in situ may be impossible without sufficient information about $(\omega, \psi)$. Moreover, if the attacker is poorly positioned---due to distance or within a high-reverberation zone---the available information admitted by $H$ may be insufficient to maintain the alignment necessary for a successful attack, regardless of the $\text{SNR}_{s}$ and $\text{SNR}_{v}$, as the attacker must operate upon the observation $z = H_{p_{s} \to p_{a}}(x)$. We believe that this formalism of relativity is crucial, as it clearly underpins the nature of risk in real world systems.

\paragraph{Acoustic Contamination} More critically, this formulation assumes $H$ is independent of $\delta$. However, in any real-time or adaptive attack, the \textbf{attacker's microphone is co-located with their own source of interference}. Consequently, the observation is not a passive window into $x$, but the mixture:
\begin{equation}
    z = H_{p_{s} \to p_{a}}(x) + H_{p_{a} \to p_{a}}(\delta)\enspace.
\end{equation}
Here $H_{p_{a} \to p_{a}}$ represents the near-field feedback of the perturbation. Because the distance $d(p_{a}, p_{a})$ is typically small, the energy of $\delta$ dominates the observation space $\mathcal{Z}$ at the attacker's sensor. This creates a \textbf{deleterious feedback loop}, in that the moment that the learned mechanism $f_{\theta}$ emits a perturbation to influence the ASR, it blinds itself to the underlying signal $x$. This self-contamination significantly reduces the effective information (and SNR) for the attacker to a level where subsequent adaptation or continued alignment with the waveform $x$ is physically impossible without sophisticated echo cancellation, which would still modify the available information.

\section{Physical Attacks, and the Fundamental Realities of Scale}\label{sec:physical_attacks}

The content above suggests that there may well be no space for digital-only attacks to provide real insights into the vulnerabilities of ASR models, as they are simply too divorced from the environments in which these models actually exist. This is not to say that physical domain attacks are a viable pathway for uncovering risks either. Consider the mechanism of a typical adversarial attack: an input $x$ is drawn from a distribution $\mathcal{D}$, and is passed through $M$ to construct a perturbation $\delta$ via gradient access. However, if we intend to consider the physical domain, this process is constrained by:%
\begin{itemize}[leftmargin=*, noitemsep, topsep=0pt]
    \item \textbf{Temporal Barriers}: Unlike image-based attacks, physical audio ingestion is lower-bounded by the utterance length. As we will discuss in subsequent sections, we consider 8 million experimental evaluations. In a physical environment, replicating this would take $925$ days, synchronized playback, and this is before we consider the labor cost of reconfiguring $\mathcal{P}$.
    \item \textbf{Complexity Explosion}: For a true physically aware acoustic attack, the generator can not simply learn the mapping $f_{\theta}(x)$, but rather $f_{\theta}(x; \omega, \psi, \mathbf{p})$, where $\omega \in \Omega$ represents the acoustic environment, $\psi \in \Psi$ the hardware infrastructure, and $\mathbf{p} \in \mathcal{P}$ the relative geometry over $\{s, v, a\}$.%
\end{itemize}
This last factor induces a prohibitive growth in sample complexity. It is not enough for a model to simply be robust against random noise. Instead, the physical domain requires robustness to be considered as a product of structured transformations and random elements. However, the number of observations required to generalize across the joint space of rooms $\Omega$, hardware configurations $\Psi$, and spatial layouts $\mathcal{P}$ is orders of magnitude larger than has been acknowledged in any prior works. 

To probe these limitations, we interrogated the performance of ASR systems in a complex L-shaped room (see Appendix~\ref{app:physical_exp} for details). Our empirical results expose a fundamental correlation collapse between digital metrics and physical outcomes. As is detailed in Table~\ref{tab:wer_snr_correlation}, while strictly digital perturbations exhibit a strong negative correlation ($r=-0.62$) between SNR and the Word Error Rate (WER), this relationship entirely dissipates in the acoustic environment ($r = -0.07$). 

\begin{table}[htbp]
    \centering
    \caption{Statistical correlations between WER and SNR for purely digital environments and acoustic environments with additive noise.}
    \label{tab:wer_snr_correlation}
    \begin{tabular}{lcc}
        \toprule
        \textbf{Condition} & \textbf{Correlation ($r$)} & \textbf{95\% CI} \\
        \midrule
        Purely Digital & $-0.62$ & $[-0.83, -0.24]$ \\
        Acoustic Noisy & $-0.07$ & $[-0.46, 0.34]$ \\
        \bottomrule
    \end{tabular}
\end{table}

We further observe that there are baseline levels of environmental degradation in acoustic environments, that rival the influence of adversarial perturbations in purely digital domains. Even without an active attacker, natural acoustic filtering pushes the \textit{Whisper-Base} model near its operational limits, with baseline WERs ranging from $69.8\%$ to $86.9\%$. Spatial layout dictates efficacy more than broadcast power; with experimental configurations where the attacker is closer to the victim yielding a mean WER of $90.2\%$, while equidistant nodes suffer significantly higher path losses without a commensurate increase in transcription failure. 

This demonstrates that models deployed in unconstrained physical spaces may already be operating under severe environmental duress, heavily skewing their susceptibility to further perturbation. The inability of raw digital-only metrics to capture these nonlinearities, combined with the impossible human cost of managing large-scale physical configurations even for the simplest $\mathcal{P}$, underscores the necessity of the high-throughput, acoustically-aligned simulation framework that we will now introduce.

\section{An Acoustic Simulation Framework}\label{sec:acoustic_simulation}

The preceding analysis establishes the fundamental tension in acoustic research: digital-only simulations are fast, and align with established workflows; while physical evaluations lack the scale necessary for statistical generalization. The impacts of this tension are only magnified for adversarial acoustics, due to the iterative mechanisms involved with adversarial attacks. To resolve this tension, we contend that researchers must adopt a middle ground that balances physical realism with high-throughput \emph{acoustically-aware} simulations. This is not to say that perfect-realism should be the goal, rather we contend that research methods in spaces with underlying physics should move as close to those true physics as available resources allow.

Modeling the propagation of adversarial perturbations in physical environments necessitates a selection between wave-based and geometrical acoustics (GA) simulations. Wave-based methods, such as Finite Difference Time Domain (FDTD), provide high-fidelity solutions to the wave equation but scale as $O(f^4)$~\citep{vorlander2008auralization}, rendering them prohibitively expensive for 16~kHz ASR signals. In contrast, GA methods like the Image Source Method (ISM) treat sound as rays, providing a computationally tractable approximation. 

One of our aims with this work was to develop a high-throughput acoustic environment that can be applied to AML problems, producing what we dub an \emph{acoustically-aligned} environment. As the backbone of this, we leverage the ISM as implemented in \textbf{PyRoomAcoustics}~\citep{scheibler2018pyroomacoustics} for its analytical, deterministic modeling of specular reflections, which yields a Room Impulse Response (RIR) that can be efficiently incorporated into differentiable network pipelines. The algorithmic details of this are described within Appendix~\ref{ref:bottlenecks}.

\begin{table}[t]
\caption{Comparison of acoustic simulation methodologies. ISM provides the necessary differentiability and throughput for large-scale adversarial optimization, producing significant alignment dividends over digital-only approximations covering computational cost, accuracy ($A(\cdot)$), gradient accessibility, implementation difficulty, and weaknesses.}
\centering
\small
\begin{tabularx}{\columnwidth}{l c c c c c >{\raggedright\arraybackslash}X}
\toprule
\textbf{Method} & \textbf{Cost} & \textbf{A(Low-$f$)} & \textbf{A(High-$f$)} & \textbf{Grad.} & \textbf{Impl.} & \textbf{Weakness} \\
\midrule
Wave-based & \textuparrow\textuparrow\textuparrow & High & High & Difficult & \textuparrow\textuparrow\textuparrow & $O(f^4)$ scaling; low iteration. \\
Ray Tracing & \textuparrow\textuparrow & Low & High & Stochastic & \textuparrow & Unstable gradients; graininess. \\
ISM & \textdownarrow & Mod. & High & \textbf{Analytic} & \textdownarrow & Specular only; no diffraction. \\
\bottomrule
\end{tabularx}
\end{table}

We note that such simulations are not the only viable pathway to incorporate the RIR into acoustic simulations. In fact, there exist several works that present databases of measured RIRs, that could be applied to these circumstances. However, these datasets are intrinsically tied to hardware idiosyncrasies, transducer non-linearity, and measurement noise~\citep{stan2002comparison}. Moreover, these datasets share a similar limitation to physical experiments, in that they lack the capacity for continuous parameterization and dynamic control, inhibiting their capacity for properly understanding and exploring the influence of acoustic information on adversarial attacks. 

Our framework is designed to generate a massive distribution of simulated environments. In doing so, we both demonstrate that it is possible to reach the levels of statistical significance in acoustic experimental evaluation, while simultaneously mitigating the risk of overfitting to the artifacts of a single setup. In essence, we trade the localized fidelity of a measured RIR for the global statistical significance of a high-throughput simulation, ensuring our findings describe fundamental properties of reverberant propagation rather than artifacts of a specific laboratory configuration or physical testing environment.

\subsection{The Knowledge Gradient: Quantifying Information Cost}

Our high-throughput capability allows us to formalize the \textbf{Knowledge Gradient}, which quantifies the Information Cost associated with environmental uncertainty. We explore how an attacker's utility degrades as their knowledge of the transfer function $H$ (and thus the RIR $h$) fades through
\begin{itemize}[leftmargin=*, noitemsep, topsep=0pt]
    \item \textbf{Naive ($\mathcal{K}_{naive}$):} Attacker assumes identity channels ($h \approx \delta_{dirac}$), acting as if in a digital domain.
    \item \textbf{Blind ($\mathcal{K}_{blind}$):} Attacker uses an EoT ensemble of $3$ rooms sampled from a distribution centered on the target room dimensions with $\pm 20 \%$ variance, representing coarse knowledge of scale.
    \item \textbf{Approximate ($\mathcal{K}_{approx}$):} Tighter $\pm 10\%$ variance but only one randomly sampled room, reflecting imperfect measurement without optimization over uncertainty.
    \item \textbf{Oracle ($\mathcal{K}_{oracle}$):} Perfect knowledge of pairwise RIRs, serving as an upper bound on performance, not a realistic attacker.
\end{itemize}

\begin{table}[h]
\caption{Comparison of Audio Adversarial Frameworks. \textbf{Dual-SNR} refers to separating Source (Stealth) and Victim (Interference) metrics. \textbf{Blind} refers to testing on unknown room geometries. \textbf{Physics} denotes if the method explicitly models room acoustics.}
\label{tab:comparison}
\centering
\small
\begin{tabular}{lccccr}
\toprule
Method & Attack & Stealth & Dual & Blind & Physics \\
& Type & Metric & SNR & Eval & Model \\
\midrule
C\&W \yrcite{carlini2018audio} & Digital & $L_p$ Norm & No & No & None \\
CmdrSong \yrcite{yuan2018commandersong} & Music & User Study & No & No & Air \\
Qin et al. \yrcite{qin2019imperceptible} & PsychoAc & Masking & No & No & Sim \\
Yakura \yrcite{yakura2018robust} & EoT & SNR & No & Yes & RIR+ \\
Devil's \yrcite{chen2020devil} & BlackBox & User Study & No & Yes & None \\
AdvPulse \yrcite{li2020advpulse} & Burst & User Study & No & Yes & Air \\
AdvReverb \yrcite{chen2023advreverb} & RIR-Gen & User Study & No & No & RIR \\
\textbf{Ours} & \textbf{Eval} & \textbf{Dual-SNR} & \textbf{Yes} & \textbf{Yes} & \textbf{RIR++} \\
\bottomrule
\end{tabular}
\end{table}

\subsection{Acoustically-Aligned Evaluation Framework}

Prior research has pivoted toward naturalistic disguises, such as \textit{AdvReverb}~\citep{chen2023advreverb}, which leverages reverberation to hide adversarial perturbations. However, we challenge the practical efficacy of this paradigm, in that while the \textit{format} of the noise is less suspicious, the \textit{energy} required to maintain alignment across a room may still render it detectable. 

As such, our framework shifts the focus from \textit{generation} to \textbf{rigorous evaluation}. Rather than proposing new attacks, we provide the methodology to understand why attacks that appear locally stealthy can be energetically visible when accounting for physics. As we discussed in Section~\ref{sec:acoustic_attacks}, the SNR does not exist at a single location, but is the product of modulated representations across the environment, giving rise to the \textbf{Dual-Form SNR}. The backbone of our evaluation involves evaluating adversarial attacks which seek to find a perturbation $\delta$ to maximize ASR loss at the victim
\begin{equation}\label{eqn:constraint}
    \max_{\delta} \mathcal{L}_{\text{ASR}}(M(H_{p_s \to p_v}(x) + H_{p_a \to p_v}(\delta)), T) \ \text{s.t.} \ \text{SNR}_{s} \geq c\enspace,
\end{equation}
where $c$ is the source-side energy budget. 

Ideally, the stealth constraint would be  be applied to the energy actually received by the source's monitoring microphone: $\| H_{p_a \to p_s}(\delta) \|_2 \leq c$. To maintain computational efficiency within the inner optimization loop, we exploit the fact that $H$ is a linear time-invariant operator defined by impulse response $h$, which allows us to employ Young's Convolution Inequality~\citep{young1912multiplication} to upper bound the received energy by $\| h \|_{1} \| \delta \|_2$. Assuming a constant channel response, we can effectively constrain the received energy by relaxing the constraint to the raw broadcast signal $\|\delta\|_2 \leq c'$. This relaxation avoids the significant computational overhead of performing differentiable convolutions within the inner loop of the optimizer, while ensuring the attack's energetic signature remains bounded relative to the carrier $x$. This formulation explicitly treats the RIR as an energetic barrier rather than a disguise, as seen in Table~\ref{tab:comparison}. We dub the difference $\Delta \text{SNR} = \text{SNR}_{s} - \text{SNR}_{v}$ the \textbf{Projection Cost}. Due to the Inverse Square Law and frequency-dependent absorption, maintaining the interference required to flip a model often necessitates a significantly louder broadcast at the source, revealing the Energetic Cost of acoustic projection.

We achieve these attacks using two core attack frameworks. The Fast Gradient Sign Method (FGSM) \citep{goodfellow2014explaining} generates $\delta$ by taking a single step in the direction of the gradient sign. This approach is generalized by its iterative refinement, Projected Gradient Descent (PGD) \citep{madry2017towards}, which traverses the loss landscape of the model $f$ to maximize error according to
\begin{equation}
\label{eq:pgd_attack}
x_{t+1} = \Pi_{x+\mathcal{S}} \left( x_t + \alpha \cdot \text{sign}(\nabla_{x_t} \mathcal{L}_{\text{ASR}}(f(x_t), T)) \right)
\end{equation}
where $x_0 = x$ an initial known signal, $\mathcal{L}_{\text{ASR}}$ is the ASR loss function (typically Cross-Entropy or CTC/Connectionist Temporal Classification), $\alpha$ is the step size, $T$ is the original transcription, and $\Pi_{x+\mathcal{S}}$ is a projection operator that ensures the adversarial example remains within both an $\epsilon$-bounded constraint set $\mathcal{S}$ around the original signal $x$ and the valid data range, enforcing the $c$ constraint. For the single-step FGSM attack, the update is simplified by setting the step size equal to the total budget ($\alpha = \epsilon$). For both attacks, we employ the untargeted variant to focus upon maximizing losses, rather than attempting to identify if targeted acoustics gradients are meaningful within acoustic environments. In the audio domain, these perturbations manifest as broadband noise that, while imperceptible, disrupts the spectral features used for transcription.

We emphasize that our contributions---which relate to the level of information available to the attacker, the relativity of SNR, and the value of acoustically-aligned evaluation environments---are not predicated upon the specific form of the attack. Changing the attack mechanism does not change the level of information accessible to the attacker, nor does it fundamentally change the limitations of the digital-only environments tested within prior works. Finally, the scale of our experiments (which we will now discuss) would make the more complex optimization loops for alternative attacks computationally prohibitive. %
A discussion of the implementation of our approach, and the computational considerations associated with it can be found in Appendix~\ref{ref:bottlenecks}.

\section{Experimental Results}\label{sec:results}

To statistically cover the long-tail variance of acoustic environments, we conduct this evaluation at massive scale, processing $200$ unique audio samples across $50$ distinct room geometries. Unlike standard digital evaluations that test a single attack configuration per sample, our framework performs a dense hierarchical sweep for every one of the $10{,}000$ sample--environment pairings: 
  \begin{itemize}[leftmargin=*, noitemsep, topsep=0pt]
      \item \textbf{Attack Generation (32 Variants):} For each sample, we generate the Cartesian product of 2 Optimizers (FGSM, PGD), 4 Target SNRs (15--45~dB, see Appendix~\ref{app:process} for details), and 4
  Knowledge Levels (Naive, Blind, Approx, Oracle).
      \item \textbf{Attack Evaluation (8 Contexts):} Each variant tests 4 Transfer Scenarios and 2 Defense States.%
  \end{itemize}
$256$ distinct interference and metric calculations were performed for every sample. Across $4$ model configurations, this resulted in over $10$ million scheduled attack evaluations ($8$ million completed, due to runtime limits), providing a statistically rigorous mapping of the acoustic attack landscape. Appendix~\ref{ref:bottlenecks} contains additional information about the process, and Appendix~\ref{app:physical_exp} demonstrates how physically repeating these simulations may take more than $925$ days. %

Our attacks are constructed under an $\ell_\infty$ constraint, where the perturbation budget $\epsilon$ is calibrated to an approximate target SNR, based upon a pre-computational analysis of $200$ room configurations. The difference between this and the SNR is deliberate: the $\ell_\infty$ constraint prevents the optimizer from concentrating energy in isolated, highly detectable bursts, while the $\ell_2$-based SNR provides a physically meaningful measure of the models perception of environmental impact and interference.%

\begin{figure*}%
    \centering
    \includegraphics[width=0.95\textwidth]{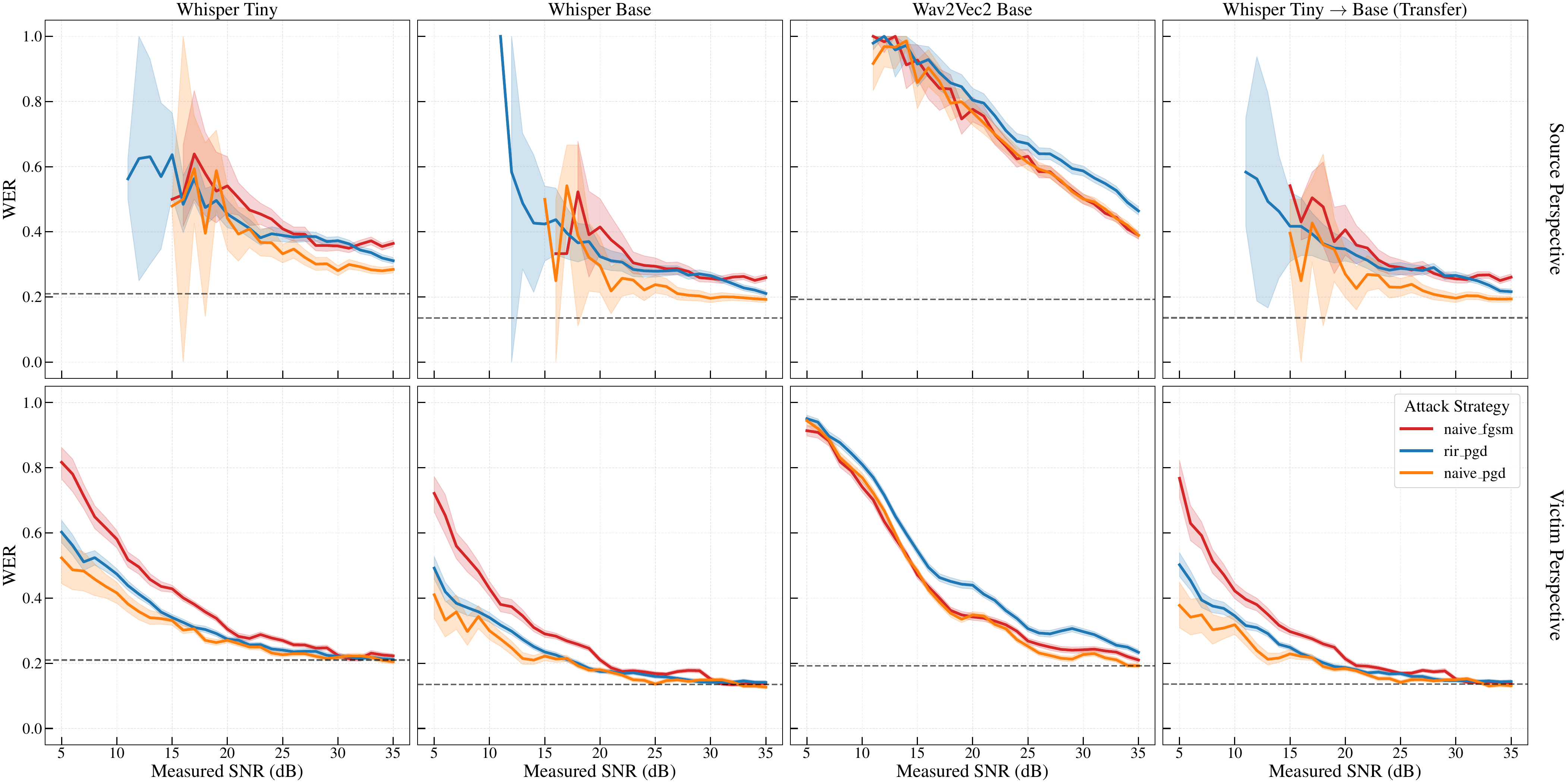}
    \caption{Gradient Misalignment across models, showing SNR at the Source (Top) or Victim (bottom) required to reach a WER. Shading represents bootstrapped $95\%$ confidence intervals.} %
    \label{fig:grad_grid}
\end{figure*}

\subsection{Gradient Misalignment and Robustness}%

Table~\ref{tab:attack_knowledge_comprehensive_table} and Figure~\ref{fig:grad_grid} reveal the counter-intuitive relationship between physical knowledge and attack efficacy. Rather than physical awareness universally improving attacks, our framework exposes stark difference in model acoustic vulnerabilities. For CTC-based models like wav2vec2, the Knowledge Gradient is effective, with the Oracle RIR approach yielding a near 100\% relative increase in the WER under attack. Conversely, for Transformer-based models like Whisper, we observe a paradox: the Naive FGSM attack (which ignores room acoustics entirely) performs comparably to, or even outperforms, attacks with perfect room knowledge. %

This one-step inverse failure is a signature of Gradient Misalignment in reverberant channels. To interrogate the drivers behind this counterintuitive phenomenon, we return to the attack mechanism. In the naive case, the $\nabla_\text{naive}$ of Equation~\ref{eq:pgd_attack} is computed on the clean model, producing a resultant perturbation that exhibits broadband phenomenology, across the entire spectral range. While the room $h$ distorts this noise, enough spectral energy survives to disrupt the ASR. 

By contrast, the RIR-aware case involves backpropagation of gradients through the RIR by way of $\nabla_\text{rir} = \nabla_\text{loss} * h_\text{reverse}$.
Crucially, real-world RIRs are characterized by deep spectral nulls, which are narrow frequency bands where the response is significantly attenuated, manifesting as destructive interference that inherently forces computational gradients to near zero. We hypothesize that these these regions significantly complicate adversarial convergence, as the $\text{sign}(\cdot)$ operation in attacks amplifies numerical noise in these nulls to $\pm \epsilon$, wasting the perturbation budget on frequencies that the room physically filters out, and directing the attack in non-productive directions. While not a total failure, RIR-FGSM provides only a marginal lift (e.g., +8\% for Whisper Base) compared to Naive FGSM (+29.9\%), confirming that one-step inversion is crippled by spectral nulls. That these results do not persist for PGD style attacks is not surprising---taking smaller steps allows PGD to optimize around nulls, while still accurately converging upon favorable solutions.

\subsection{The Acoustic Tax and Projection Cost}

\begin{table*}
\caption{Word Error Rates for a range of experimental configurations, relative to a per-model baseline. Here Transfer represents Whisper Tiny $\rightarrow$ Base (Transfer).}
\label{tab:attack_knowledge_comprehensive_table}
\centering
\small
\resizebox{\columnwidth}{!}{%
\begin{tabular}{llccccc}
\toprule
 &  & Baseline & Naive (Identity) & Blind (Dist.) & Approximate & Oracle (Known) \\
\midrule
\multirow[t]{2}{*}{\textbf{Whisper Tiny}} & FGSM & 0.21 & 0.264 (+25.7\%) & 0.226 (+7.8\%) & 0.226 (+7.8\%) & 0.226 (+7.8\%) \\
 & PGD & 0.21 & 0.239 (+13.8\%) & 0.256 (+22.1\%) & 0.257 (+22.1\%) & 0.265 (+26.0\%) \\
\cmidrule{1-7}
\multirow[t]{2}{*}{\textbf{Whisper Base}} & FGSM & 0.135 & 0.176 (+29.9\%) & 0.146 (+8.0\%) & 0.146 (+8.0\%) & 0.146 (+8.1\%) \\
 & PGD & 0.135 & 0.155 (+14.8\%) & 0.17 (+25.9\%) & 0.17 (+26.0\%) & 0.176 (+30.4\%) \\
\cmidrule{1-7}
\multirow[t]{2}{*}{\textbf{wav2vec Base}} & FGSM & 0.193 & 0.328 (+70.3\%) & 0.24 (+24.5\%) & 0.24 (+24.4\%) & 0.24 (+24.4\%) \\
 & PGD & 0.193 & 0.325 (+69.0\%) & 0.329 (+71.0\%) & 0.326 (+69.4\%) & 0.375 (+94.5\%) \\
\cmidrule{1-7}
\multirow[t]{2}{*}{\textbf{Transfer}} & FGSM & 0.136 & 0.178 (+30.8\%) & 0.148 (+8.7\%) & 0.148 (+8.7\%) & 0.148 (+8.8\%) \\
 & PGD & 0.136 & 0.157 (+15.4\%) & 0.172 (+26.6\%) & 0.173 (+26.9\%) & 0.179 (+31.7\%) \\
\bottomrule
\end{tabular}
}
\end{table*}

\begin{wrapfigure}{r}{0.4\textwidth}
\vspace{-25pt} 
\includegraphics[width=.4\textwidth]{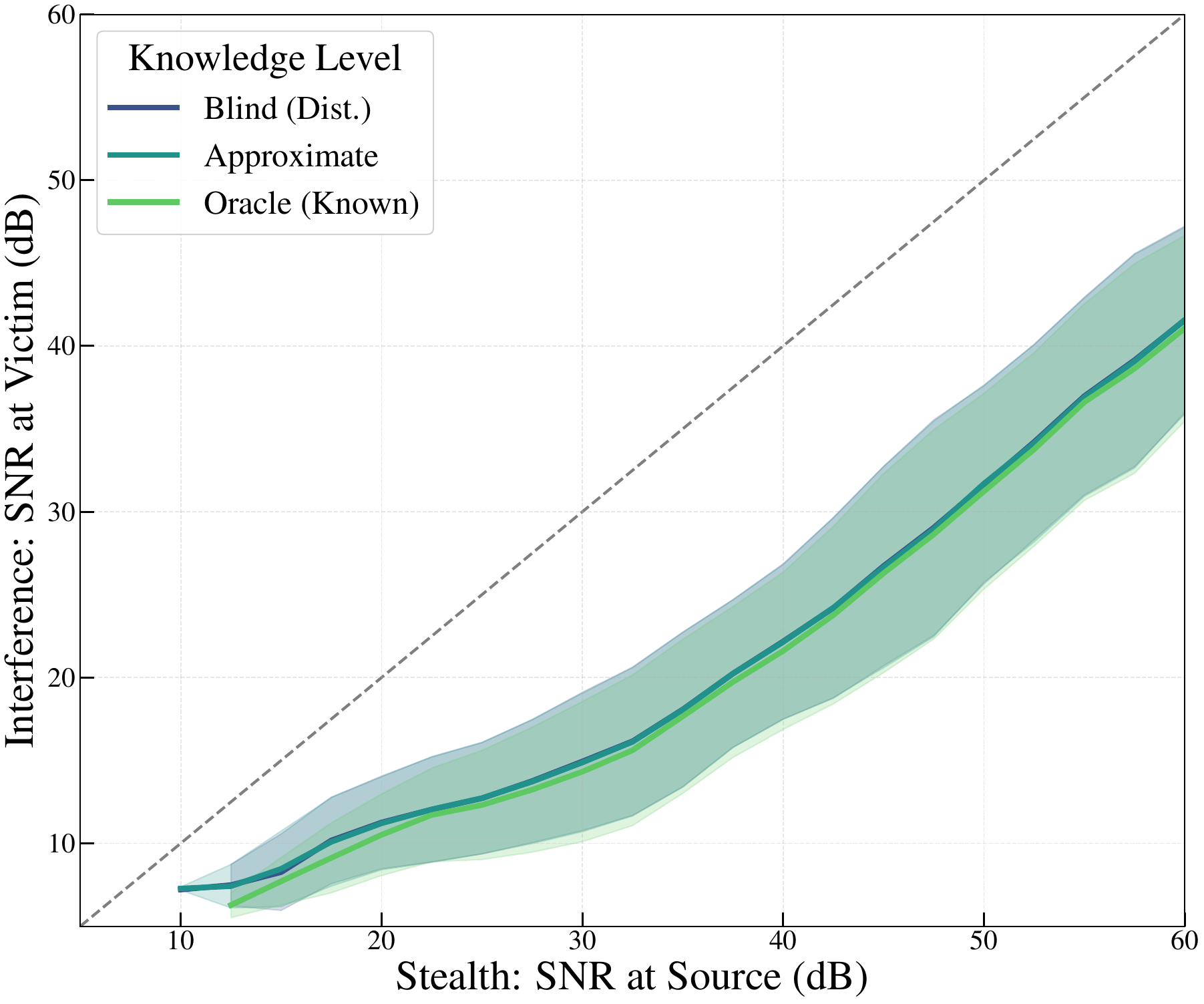}%
      \caption{Consolidated Projection Cost: divergence from the identity line (dashed) quantifies the Acoustic Tax inherent in OTA projection across
  tested room geometries. Shading: 1$\sigma$ variance.}
      \label{fig:proj_consolidated}
    \vspace{-35pt} 
\end{wrapfigure}

Figure \ref{fig:proj_consolidated} visualizes the inherent decoupling of Source and Victim SNR, with our analysis revealing a significant perceptual disconnect between the two locations: an attack that achieves a successful interference level of $\text{SNR}_\text{victim} \approx 20$~dB at the target often manifests as a highly stealthy $\text{SNR}_\text{source} \approx 40$~dB at the source monitoring microphone. This is reinforced by Figure~\ref{fig:geometry_of_WER} demonstrating that WER is primarily correlated with victim SNR. This strongly suggests that future geometrically aware attacks---particularly those using directional acoustic devices---could further exploit this decoupling to yield attacks that are \textbf{both stealthy and efficacious}. 

The heatmap in Figure \ref{fig:tax_heatmap} quantifies this relationship, demonstrating that the gap between victim interference and source audibility is not a constant but a function of room geometry and reverberation. This result provides a critical physical nuance to prior works on perceptual stealth \citep{szurley2019perceptual}; \textbf{while an attack may be masked psychoacoustically at one location, its energetic signature is non-uniform, with a differential of over $30$~dB for large, reverberant spaces} $\mathbf{RT_{60} > 0.6}$ \textbf{s)}. %
This implies that the carrier device's own monitoring system is a poor proxy for model impact, %
and conversely, \textbf{that reporting a single SNR scalar is a physically incomplete measure of an attack's presence}. This spatial invariance shown in Figure~\ref{fig:dist_grid} further reinforces the disconnect: for an attacker increasing broadcast power to compensate for distance ($d_{AV}$), the impact on the model remains consistent, confirming that OTA attacks are limited not by placement, but by the attacker's ability to balance minimizing energetic visibility with attack utility. 

\subsection{Model Architecture and Transferability}

As quantified in the Oracle RIR column of Table \ref{tab:attack_knowledge_comprehensive_table}, Wav2Vec2 (CTC-based) proved significantly more brittle to reverberant interference than Whisper (Transformer-based), showing a 94.5\% relative increase in WER compared to Whisper Base's 30.4\%. Adversarial examples generated on Whisper-Tiny transferred effectively to Whisper-Base, maintaining a consistent 31.7\% lift over baseline, suggesting a robustness against model scale.%

The impact of these attacks is further contextualized by discussions regarding adversarial defenses (Appendix~\ref{app:defenses}), the impact of computational cost (Appendix~\ref{ref:bottlenecks}), and physical implications of acoustic modeling, including the Acoustic Tax (Figure \ref{fig:tax_heatmap}), the preservation of signal quality (Figure \ref{fig:quality_attack}), and the impact of spatial properties (Appendix~\ref{app:physical}). %

\section{Conclusion}

By integrating adversarial attacks into a high-throughput acoustic simulation, we have elucidated the impact of two fundamental properties of acoustic AML: the Knowledge Gradient and SNR relativity. Our analysis reveals a crucial paradox in OTA research, in that by abstracting away the acoustic environment we may have both overestimated realisable stealth, and underestimated the brute-force simplicity required to influence model predictions. 

More broadly, this work demonstrates how %
a physics-based lens reveals problems that are obscured within digital-only environments. We demonstrate that digital-only insights do not necessarily translate to physical simulations; specifically, the energetic disconnect between emissions and reception at the source and victim creates an acoustic tax that dictates the feasibility of any real-world attack. Our results suggest that while an attack may be masked psychoacoustically at one location, its energetic signature is non-uniform across the acoustic space, rendering single-scalar SNR reports an incomplete and potentially misleading measure of an attack's presence. 

We emphasize that our framework does not claim to represent a perfect acoustic digital twin. Rather, we contend that for research in modalities with underlying physical constraints, investigators must move as close as possible to that physical reality as available resources allow, without compromising on statistical viability. Our 8 million tests (which would take $925$ days to replicate) embrace, rather than abstract, the environment, and all us to demonstrate that aggregate performance across a statistical ensemble identifies fundamental vulnerabilities rather than measurement artifacts. We provide the metrological foundations for a new era of systemic robustness evaluation, moving the field beyond mathematical convenience toward a deeper understanding of the risks facing deployed ASR.%

\FloatBarrier

\newpage

\section*{Impact Statement}

This work explores how simulating acoustically-aligned environments can provide a more realistic framework for assessing risk in acoustic AML. While the overall aim is to improve model robustness by increasing the alignment between AML and acoustic physics, we would be remiss to not discuss the fact that this work takes a primarily adversarial lens to test current behaviors. 

In general, adversarial attacks are perceived as having the potential to cause harm, due to their focus upon manipulating real world systems. With this said, there is also the potential for benefits to be induced---for example, in the face of increasingly normalized surveillance, it may well be the case that adversarial attacks may induce privacy, creating a net public good. More broadly, we would argue that current offensive and defensive mechanisms within this space fail from a crucial alignment failure, that renders security research in this space as security theatre. 

Being able to bridge the gap between digital domain acoustic attacks (which are favorable for their throughput) and OTA attacks is of critical importance for security researchers to truly understand the vulnerabilities in ASR systems. While our primary focus in this work was an attack-based application, our framework sets the foundations for rigorous, acoustically-aligned experimental research into defensive mechanisms.

\bibliography{references}

@inproceedings{abdullah2021sok,
    author = {Hadi Abdullah and Kevin Warren and Vincent Bindschaedler and Nicolas Papernot and Patrick Traynor},
    booktitle = {IEEE Symposium on Security and Privacy (SP)},
    pages = {730-747},
    title = {{{S}o{K}: {T}he {F}aults in our {A}{S}{R}s: {A}n {O}verview of {A}ttacks against {A}utomatic {S}peech {R}ecognition and {S}peaker {I}dentification {S}ystems}},
    year = {2021}
}

@inproceedings{athalye2018synthesizing,
    author = {Athalye, Anish and Engstrom, Logan and Ilyas, Andrew and Kwok, Kevin},
    booktitle = {International Conference on Machine Learning (ICML)},
    organization = {PMLR},
    pages = {284--293},
    title = {{S}ynthesizing {R}obust {A}dversarial {E}xamples},
    year = {2018}
}

@article{baevski2020wav2vec,
    author = {Baevski, Alexei and Zhou, Yuhao and Mohamed, Abdelrahman and Auli, Michael},
    journal = {Advances in Neural Information Processing Systems (NeurIPS)},
    pages = {12449--12460},
    title = {wav2vec 2.0: {A} {F}ramework for {S}elf-{S}upervised {L}earning of {S}peech {R}epresentations},
    volume = {33},
    year = {2020}
}

@inproceedings{carlini2018audio,
    author = {Carlini, Nicholas and Wagner, David},
    booktitle = {2018 IEEE Security and Privacy Workshops (SPW)},
    number = {},
    pages = {1-7},
    title = {{A}udio {A}dversarial {E}xamples: {T}argeted {A}ttacks on {S}peech-to-{T}ext},
    volume = {},
    year = {2018}
}

@inproceedings{chen2020devil,
    author = {Chen, Yuxuan and Yuan, Xuejing and Zhang, Jiangshan and Zhao, Yue and Zhang, Shengzhi and Chen, Kai and Wang, XiaoFeng},
    booktitle = {29th USENIX Security Symposium (USENIX Security 20)},
    pages = {2667--2684},
    title = {{{D}evil{\textquoteright}s} {W}hisper: {A} {G}eneral {A}pproach for {P}hysical {A}dversarial {A}ttacks against {C}ommercial {B}lack-box {S}peech {R}ecognition {D}evices},
    year = {2020}
}

@article{chen2023advreverb,
    author = {Chen, Meng and Lu, Li and Yu, Jiadi and Ba, Zhongjie and Lin, Feng and Ren, Kui},
    journal = {IEEE Transactions on Information Forensics and Security},
    number = {},
    pages = {1948-1962},
    title = {{A}dv{R}everb: {R}ethinking the {S}tealthiness of {A}udio {A}dversarial {E}xamples to {H}uman {P}erception},
    volume = {19},
    year = {2024}
}

@inproceedings{goodfellow2014explaining,
    author = {Ian J. Goodfellow and
Jonathon Shlens and
Christian Szegedy},
    booktitle = {International Conference on Learning Representations, {ICLR}},
    title = {{E}xplaining and {H}arnessing {A}dversarial {E}xamples},
    volume = {3},
    year = {2015}
}

@inproceedings{li2020advpulse,
    author = {Li, Zhuohang and Wu, Yi and Liu, Jian and Chen, Yingying and Yuan, Bo},
    booktitle = {Proceedings of the 2020 ACM SIGSAC Conference on Computer and Communications Security},
    numpages = {14},
    pages = {1121–1134},
    series = {CCS '20},
    title = {{A}dv{P}ulse: {U}niversal, {S}ynchronization-free, and {T}argeted {A}udio {A}dversarial {A}ttacks via {S}ubsecond {P}erturbations},
    year = {2020}
}

@inproceedings{madry2017towards,
    author = {Aleksander Madry and Aleksandar Makelov and Ludwig Schmidt and Dimitris Tsipras and Adrian Vladu},
    booktitle = {International Conference on Learning Representations (ICLR)},
    title = {{T}owards {D}eep {L}earning {M}odels {R}esistant to {A}dversarial {A}ttacks},
    volume = {6},
    year = {2018}
}

@inproceedings{olivier2022there,
    author = {Olivier, Raphael and Raj, Bhiksha},
    booktitle = {INTERSPEECH},
    title = {{{T}here is {M}ore {T}han {O}ne {K}ind of {R}obustness: {F}ooling {W}hisper with {A}dversarial {E}xamples}},
    year = {2023}
}

@inproceedings{panayotov2015librispeech,
    author = {Panayotov, Vassil and Chen, Guoguo and Povey, Daniel and Khudanpur, Sanjeev},
    booktitle = {IEEE International Conference on Acoustics, Speech and Signal Processing (ICASSP)},
    organization = {IEEE},
    pages = {5206--5210},
    title = {{L}ibrispeech: {A}n {{A}{S}{R}} corpus based on public domain audio books},
    year = {2015}
}

@inproceedings{qin2019imperceptible,
    author = {Qin, Yao and Carlini, Nicholas and Goodfellow, Ian and Cottrell, Garrison and Raffel, Colin},
    booktitle = {International Conference on Machine Learning (ICML)},
    title = {{I}mperceptible, {R}obust, and {T}argeted {A}dversarial {E}xamples for {A}utomatic {S}peech {R}ecognition},
    year = {2019}
}

@inproceedings{radford2023robust,
    author = {Radford, Alec and Kim, Jong Wook and Xu, Tao and Brockman, Greg and McLeavey, Christine and Sutskever, Ilya},
    booktitle = {International Conference on Machine Learning (ICML)},
    organization = {PMLR},
    pages = {28448--28467},
    title = {{R}obust {S}peech {R}ecognition via {L}arge-{S}cale {W}eak {S}upervision},
    volume = {40},
    year = {2023}
}

@inproceedings{scheibler2018pyroomacoustics,
    author = {Scheibler, Robin and Bezzam, Eric and Dokmani{\'c}, Ivan},
    booktitle = {2018 IEEE International Conference on Acoustics, Speech and Signal Processing (ICASSP)},
    organization = {IEEE},
    pages = {351--355},
    title = {{P}yroomacoustics: {A} {{P}ython} package for audio room simulation and array processing algorithms},
    year = {2018}
}

@article{stan2002comparison,
    author = {Stan, Guy-Bart and Embrechts, Jean Jacques and Archambeau, Dominique},
    journal = {Journal of the Audio Engineering Society},
    month = {04},
    pages = {249-262},
    title = {{C}omparison of {D}ifferent {I}mpulse {R}esponse {M}easurement {T}echniques},
    volume = {50},
    year = {2002}
}

@article{szurley2019perceptual,
    author = {Szurley, Joseph and Kolter, J Zico},
    journal = {arXiv preprint arXiv:1906.06355},
    title = {{P}erceptual {B}ased {A}dversarial {A}udio {A}ttacks},
    year = {2019}
}

@book{vorlander2008auralization,
    author = {Vorl{\"a}nder, Michael},
    publisher = {Springer},
    title = {{A}uralization: {F}undamentals of {A}coustics, {M}odelling, {S}imulation, {A}lgorithms and {A}coustic {V}irtual {R}eality},
    year = {2008}
}

@inproceedings{yakura2018robust,
    author = {Yakura, Hiromu and Sakuma, Jun},
    booktitle = {Proceedings of the Twenty-Eighth International Joint Conference on
Artificial Intelligence, {IJCAI-19}},
    doi = {10.24963/ijcai.2019/741},
    month = {7},
    pages = {5334--5341},
    publisher = {International Joint Conferences on Artificial Intelligence Organization},
    title = {{R}obust {A}udio {A}dversarial {E}xample for a {P}hysical {A}ttack},
    year = {2019}
}

@inproceedings{yuan2018commandersong,
    author = {Yuan, Xuejing and Chen, Yuxuan and Zhao, Yue and Long, Yunhui and Liu, Xiaokang and Chen, Kai and Zhang, Shengzhi and Huang, Heqing and Wang, Xiaofeng and Gunter, Carl A},
    booktitle = {USENIX Security Symposium},
    title = {{{C}ommander{S}ong}: {A} {S}ystematic {A}pproach for {P}ractical {A}dversarial {V}oice {R}ecognition},
    year = {2018}
}

@inproceedings{zhang2024advsv,
    author = {Wang, Li and Li, Jiaqi and Luo, Yuhao and Zheng, Jiahao and Wang, Lei and Li, Hao and Xu, Ke and Fang, Chengfang and Shi, Jie and Wu, Zhizheng},
    booktitle = {IEEE International Conference on Acoustics, Speech and Signal Processing (ICASSP)},
    organization = {IEEE},
    pages = {4555-4559},
    title = {{A}dv{S}{V}: {A}n {O}ver-the-{A}ir {A}dversarial {A}ttack {D}ataset for {S}peaker {V}erification},
    year = {2024}
}

@article{young1912multiplication,
  title={On the {M}ultiplication of {S}uccessions of {F}ourier {C}onstants},
  author={Young, William Henry},
  journal={Proceedings of the Royal Society of London. Series A, Containing Papers of a Mathematical and Physical Character},
  volume={87},
  number={596},
  pages={331--339},
  year={1912},
  publisher={The Royal Society London}
}

@article{draper2025clinical,
  title    = {Clinical {AI} {S}cribes in {P}rimary {C}are: {A}ccuracy, {E}rror {S}everity and {I}mplications for {C}linical {P}ractice},
  author   = {Draper, Thomas C and Cox, Timothy and Lamb-Riddell, Kathryn and Moretti, Luigi Andrea and McCormick, John and Trowell, Stephen and Kiely, Janice and Luxton, Richard},
  journal  = {BMJ Digital Health \& AI},
  volume   =  {1},
  number   =  {1},
  pages    = {e000092},
  month    =  sep,
  year     =  {2025},
  doi = {10.1136/bmjdhai-2025-000092}
}

@inproceedings {zhang2024laseradv,
author = {Guoming Zhang and Xiaohui Ma and Huiting Zhang and Zhijie Xiang and Xiaoyu Ji and Yanni Yang and Xiuzhen Cheng and Pengfei Hu},
title = {{LaserAdv}: Laser {A}dversarial {A}ttacks on {S}peech {R}ecognition {S}ystems},
booktitle = {33rd USENIX Security Symposium (USENIX Security 24)},
year = {2024},
isbn = {978-1-939133-44-1},
address = {Philadelphia, PA},
pages = {3945--3961},
publisher = {USENIX Association},
}

@inproceedings{carlini2017towards,
  title={Towards {E}valuating the {R}obustness of {N}eural {N}etworks},
  author={Carlini, Nicholas and Wagner, David},
  booktitle={2017 {IEEE} {S}ymposium on {S}ecurity and {P}rivacy ({SP})},
  pages={39--57},
  year={2017},
  organization={IEEE}
}

@inproceedings{schonherr2019psychoacoustic_hiding,
  author    = {Sch{\"o}nherr, Lea and Kohls, Katharina and Zeiler, Steffen and Holz, Thorsten and Kolossa, Dorothea},
  title     = {Adversarial {A}ttacks {A}gainst {A}utomatic {S}peech {R}ecognition {S}ystems via {P}sychoacoustic {H}iding},
  booktitle = {Network and Distributed System Security Symposium ({NDSS})},
  year      = {2019},
}

@article{sun2024commanderuap,
  author  = {Sun, Qibin and Chen, Shun and Zhai, Yingbin and Liu, Yang and Zhong, Zhisheng},
  title   = {{CommanderUAP}: {P}ractical and {T}ransferable {U}niversal {A}dversarial {P}erturbations on {ASR} {S}ystems},
  journal = {Cybersecurity},
  year    = {2024},
}
\bibliographystyle{icml2026}

\appendix

\section{Physical Validation of Concepts}~\label{app:physical_exp}

As Section~\ref{sec:physical_attacks} discussed, practical physical validation of adversarial attacks very quickly becomes nigh-on impossible, when attempting to consider the physical environment. Consider for example a PGD style attack that is run for $n$ iterations on $s$ samples that is $t$ seconds long, over $v$ different physical configurations. Even a model that can process samples faster than the time it takes to listen to them still must listen to the sample first, so, if $r$ is the real-time-factor associated with processing, naively, the computational time associated with an attack $\tau$ is strictly $\tau > (1 + r) \times n \times s \times t \times v$. 

Complicating this is the need for the signal environment to be quiet at the beginning of each attack, which requires that there is a delay between iterations of the attack, and the signals at each of the three node locations (source, attacker, victim) must be synchronized. Moreover, each of the $v$ different physical configurations requires physical intervention to manually change the configuration. 

When sample lengths are on the order of $30$ seconds, a PGD attack involving $50$ steps (which may be ultimately conservative due to the nature of acoustic hulls) that requires $5$ seconds of silence to reset the acoustic environment will require $\tau > 30 (1 + r)$ minutes to attack a single sample, in a single room configuration. Complicating things further is that no realistic  environment is acoustically isolated, and as such there is a need to repeat attacks against samples multiple times to ensure rigor, producing a further multiplicative scaling for the costs and associated times. 

As such, conducting enough experiments to achieve statistically significant results under such a framework is practically infeasible when considering time and labor costs. This observation underscores the utility of our acoustically-aligned simulations for exploring the robustness and security properties of models that are deployed in the real-world, without requiring the time associated with real-world simulations. 

Rather than attempting to validate our $8$ million physical experiments directly, we considered two different perspectives. The first was to leverage an externally-validated modeling framework for producing RIRs, as was discussed within Section~\ref{sec:acoustic_simulation}. 

The second validation stage was to consider a physical experiment using an over-the-air attack with our three nodes positioned within a complex L-shaped room with maximal dimensions of $5$m $\times 6$m, with furniture, open windows, and open doors. Within this environment, rather than testing PGD or FGSM style attacks, we considered strictly the influence of bounded white noise, to limit the temporal utilization of the room environment. 

Within this acoustic space, we positioned our three nodes in six different physical configurations, detailed as follows:
\begin{itemize}[leftmargin=*, noitemsep, topsep=0pt]
    \item \textbf{Scenario 1}: The nodes are positioned in the ordering of Victim, Source, Attacker, with a spacing of $1$ m between nodes.
    \item \textbf{Scenario 2}: Same as Scenario 1, however with a spacing of $3$ m between nodes.
    \item \textbf{Scenario 3}: The nodes are positioned in the ordering of Victim, Attacker, Source with a spacing of $1$ m between nodes.
    \item \textbf{Scenario 4}: Same as Scenario 3, however with a spacing of $3$ m between nodes.
    \item \textbf{Scenario 5}: The nodes are positioned in an equidistant manner, with a spacing of $1$ m between nodes.
    \item \textbf{Scenario 6}: Same as Scenario 5, however with a spacing of $3$ m between nodes.
\end{itemize}

Over these six different physical configurations, we executed a factorial experimental design across three independent variables: noise magnitude,  source-mixing mode, and signal content.

To ensure statistical representativeness and mitigate speaker-dependent bias, we utilized a fixed corpus of $10$ randomly sampled utterances from the LibriSpeech \textit{clean-100} dataset. For each spatial configuration, we evaluated two distinct operational modes for the attacker. In \textbf{Mode 1 (Remote Interference)}, the source node broadcast clean speech while the attacker node broadcast bounded Gaussian white noise from a distinct spatial coordinate. In \textbf{Mode 2 (Local Interference)}, the attacker noise was computationally mixed with the speech signal and broadcast solely from the source node, while the remote attacker node   remained silent. This allows for an empirical comparison between local signal corruption and remote acoustic masking.

The noise intensity was varied across three magnitudes ($\epsilon \in \{0.0, 0.5, 1.0\}$), where $\epsilon=1.0$ represents a peak amplitude normalized to the signal's maximum digital headroom. To ensure temporal synchronization across the distributed nodes, each trial was preceded and followed by a $5,000$ Hz synchronization pulse (50ms duration) separated by $100$ms silence buffers. This enabled precise post-hoc temporal alignment via cross-correlation, allowing us to isolate the content signal and factor out network-induced trigger jitter.

For the final evaluation, we utilized the \textit{Whisper-Base} transformer-based ASR model. We computed the Word Error Rate (WER) by comparing the victim's captured transcription against a baseline transcription of the digital source files. Simultaneously, we quantified the acoustic channel degradation using two metrics: \textit{Absolute SNR}, calculated relative to the pristine digital reference to measure total path loss and reverberation, and \textit{Relative SNR}, calculated against the $\epsilon=0.0$ baseline recording to isolate the specific masking efficacy of the added white noise.

In total, each physical session comprised $360$ trials ($10 \text{ samples} \times 3 \text{ magnitudes} \times 2 \text{ modes} \times 6 \text { physical configurations}$), producing $1080$ distinct physical recordings for every test. To ensure performance, we then repeated each of these tests $3$ times. To prevent hardware-level automatic gain control (AGC) or non-linear compression from contaminating the results, the source playback was scaled to a constant $0.75$ of the digital maximum. 

That each trial required more than $5$ hours ($15$ hours when factoring in the $3$ repetitions) of time clearly demonstrates the futility of large scale experimental evaluations, and the critical importance of studying acoustics in acoustically-aligned environments, where real-world physics factor into experimental outcomes. Our high-throughput simulation framework is vital for scaling acoustic AML research beyond toy examples. Replicating the breadth of our experimental results would not be possible under real world conditions.

This is especially so when these scenarios required constant human engagement to change physical configurations, and monitor for intrusive external noises that required re-running a particular experiment. And, crucially, these experiments are not adaptive---incorporating a multi-step attack framework would lead to a multiplicative increase in the associated times and costs. 

\subsection{Experimental Results}

Our physical validation experiments yield several critical insights that challenge prevailing assumptions in adversarial machine learning for acoustic systems, highlighting the stark contrast between digital simulations and physical realities.

\paragraph{The Disconnect Between Digital SNR and Physical Transcriptions}Most notably, the data reveals a fundamental collapse in the correlation between Signal-to-Noise Ratio (SNR) and Word Error Rate (WER) when transitioning from digital to physical environments. As detailed in Table~\ref{tab:wer_snr_correlation}, purely digital perturbations exhibit a strong negative correlation ($r = -0.62$), suggesting that as SNR decreases, WER predictably increases. However, in physical, noisy acoustic environments, this correlation entirely dissipates ($r = -0.07$). This empirically demonstrates that evaluating acoustic adversarial attacks relying solely on digital-environments is methodologically flawed; real-world variables such as multipath propagation, room impulse response (RIR), and destructive interference exert a non-linear influence on model transcription accuracy that raw digital-only metrics cannot capture. This breaks the implicit assumption in past works that digital only SNR and WER would be meaningfully correlated, and aligns with our acoustically-aligned numerical experiments.

\paragraph{Baseline Environmental Degradation} These results clearly demonstrate that the physical space acts as a formidable adversary to acoustic models, underpinning the utility for acoustic-awareness for both attacks and general model performance. Even in the absence of an active attacker (Magnitude $0.0$), the baseline WER for the Whisper-Base model ranges from $69.8 \%$ to $86.9\%$ in Table~\ref{tab:snr_wer_scenarios}. The acoustic filtering of our physical testing environment, combined with standard physical obstacles, pushes the ASR model near its operational limits, even before any malicious payload is introduced. This highlights a critical, inherent vulnerability: models deployed in unconstrained physical spaces are already operating under severe environmental duress, heavily skewing their susceptibility to further perturbation. 

\paragraph{Spatial Dynamics and the Non-Linear Nature of Vulnerabilities} Unsurprisingly, the physical layout of the nodes heavily dictates the efficacy of the attack, although the actual results do not necessarily fully align with mathematical intuition. Table~\ref{tab:spatial_layout_summary} illustrates that configuring the attacker closer to the victim yields the highest mean WER at $90.2\%$, as the attacker's broadcast suffers less environmental attenuation than the target speech. Interestingly, the equidistant configurations produce the most severe Mean Absolute SNR ($-30.9$ dB), but a marginally lower Mean WER ($89.3\%$). This anomaly further corroborates the findings from Table~\ref{tab:wer_snr_correlation}, in that severe signal degradation does not strictly guarantee a corresponding maximization of transcription failure, when the acoustic signal is exposed to physical attenuation.

\begin{table}[htbp]
    \centering
    \caption{SNR, WER, and estimated environmental loss across different spatial configurations and noise scales. Relative SNR compares the heard signal at the victim without noise versus with noise. Absolute SNR compares the heard signal at the victim against the original digital signal. Scenarios 1--2: source is closer to the victim; Scenarios 3--4: attacker is closer; Scenarios 5--6: equidistant. Odd numbered scenarios denote 1m spacing between devices, while even scenarios denote 3m spacing.}
    \label{tab:snr_wer_scenarios}
    \begin{tabular}{clccc}
        \toprule
        \multirow{2}{*}{\textbf{Scenario}} & \multirow{2}{*}{\textbf{Metric}} & \multicolumn{3}{c}{\textbf{Noise Magnitude}} \\
        \cmidrule(lr){3-5}
        & & \textbf{Mag 0.0} & \textbf{Mag 0.5} & \textbf{Mag 1.0} \\
        \midrule
        \multirow{3}{*}{1} 
        & Abs SNR & $-12.5$ & $-16.2$ & $-16.3$ \\
        & Rel SNR & $\infty$ & $6.1$ & $-13.8$ \\
        & WER & 69.8\% & 85.5\% & 91.3\% \\
        \midrule
        \multirow{3}{*}{2} 
        & Abs SNR & $-29.6$ & $-41.7$ & $-38.7$ \\
        & Rel SNR & $\infty$ & $-45.1$ & $-47.9$ \\
        & WER & 86.9\% & 87.6\% & 94.2\% \\
        \midrule
        \multirow{3}{*}{3} 
        & Abs SNR & $-34.0$ & $-30.3$ & $-29.6$ \\
        & Rel SNR & $59.8$ & $-47.4$ & $-65.1$ \\
        & WER & 78.7\% & 105.8\% & 94.1\% \\
        \midrule
        \multirow{3}{*}{4} 
        & Abs SNR & $-29.3$ & $-31.1$ & $-27.7$ \\
        & Rel SNR & $\infty$ & $-3.8$ & $-21.6$ \\
        & WER & 75.9\% & 87.9\% & 99.7\% \\
        \midrule
        \multirow{3}{*}{5} 
        & Abs SNR & $-35.1$ & $-38.5$ & $-33.9$ \\
        & Rel SNR & $\infty$ & $-9.6$ & $-2.2$ \\
        & WER & 75.5\% & 92.5\% & 96.2\% \\
        \midrule
        \multirow{3}{*}{6} 
        & Abs SNR & $-23.2$ & $-27.4$ & $-28.4$ \\
        & Rel SNR & $\infty$ & $-27.8$ & $-16.7$ \\
        & WER & 81.4\% & 95.2\% & 97.2\% \\
        \bottomrule
    \end{tabular}
\end{table}

\paragraph{Distance Penalties and Remove Masking Efficacy} Finally, these experiments highlight the harsh realities of the inverse-square law upon practical attacks, and the crucial importance of physical configuration in determining attack  performance. Increasing the node spacing from $1$ to $3$ meters incurs severe distance penalties. For instance in Scenario 3, the Magnitude $0.5$ case produces a WER of $105.8\%$, indicating severe model hallucinations. However, extending the distance to 3 meters (Scenario 4) allows the room to diffuse the impact of this effect, capping WER at $87.9\%$. Despite these attentuation penalties, the precipitous drops in Relative SNR across the active attack scenarios proves that remote acoustic masking can aggressively overpower the source signal at the target microphone. This validates over-the-air remote interference as a highly potent attack vector, independent of local signal corruption. 

\begin{table}[htbp]
    \centering
    \caption{Mean Absolute SNR and Mean WER across varying spatial layouts.}
    \label{tab:spatial_layout_summary}
    \begin{tabular}{lcc}
        \toprule
        \textbf{Spatial Layout} & \textbf{Mean Abs SNR} & \textbf{Mean WER} \\
        \midrule
        Source Closer (S1, S2)    & $-25.2$ dB & 85.7\% \\
        Attacker Closer (S3, S4)  & $-30.4$ dB & 90.2\% \\
        Equidistant (S5, S6)      & $-30.9$ dB & 89.3\% \\
        \bottomrule
    \end{tabular}
\end{table}

\section{Physical Implications of Acoustic Modeling}\label{app:physical}

\begin{figure*}%
      \centering
      \includegraphics[width=0.65\textwidth]{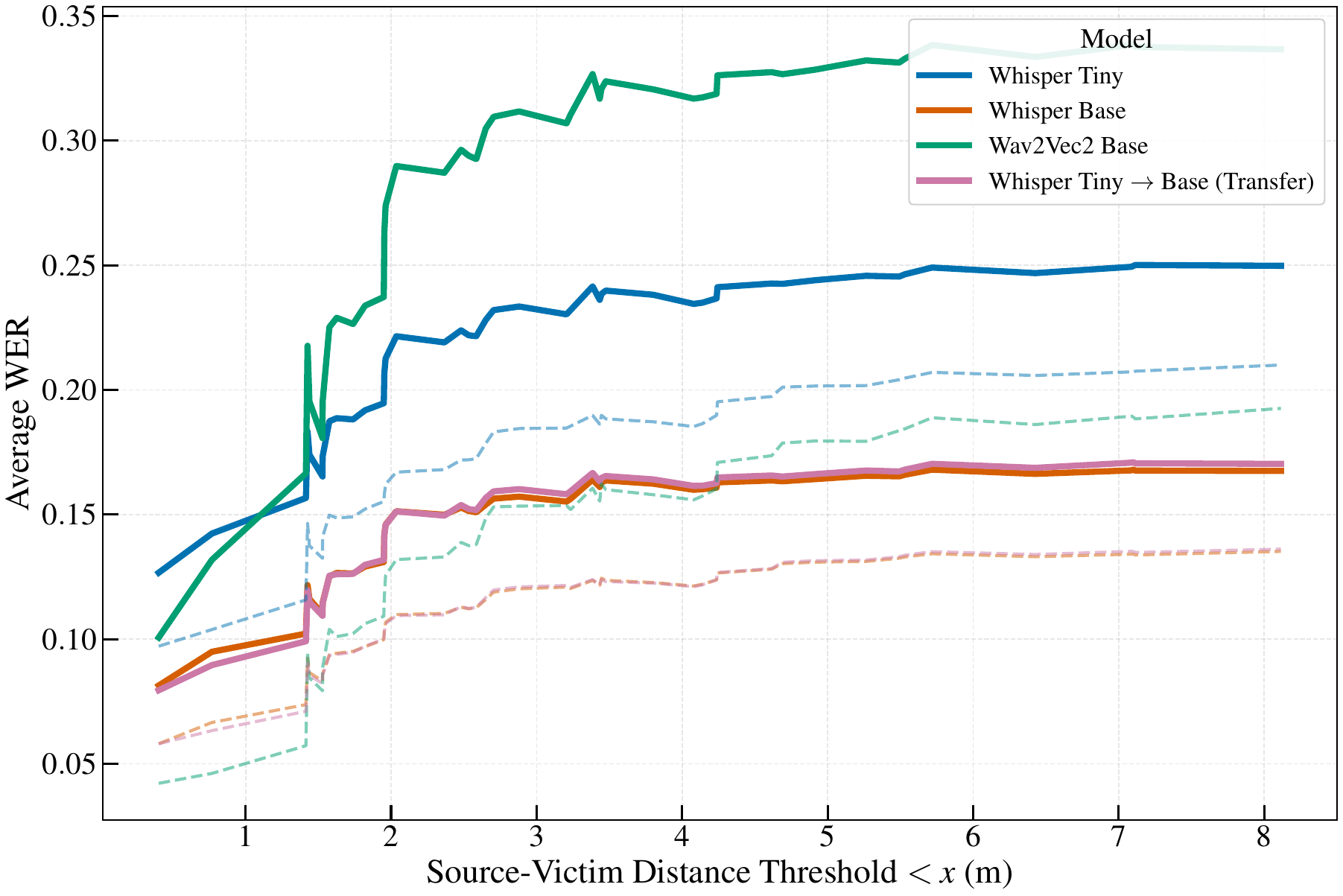}
      \caption{Relationship between average WER and Source-Victim Distance. Attacked performance is represented by solid lines, clean performance by dashed lines. }
      \label{fig:SV_Distance}
  \end{figure*}

\begin{figure*}%
      \centering
      \includegraphics[width=0.85\textwidth]{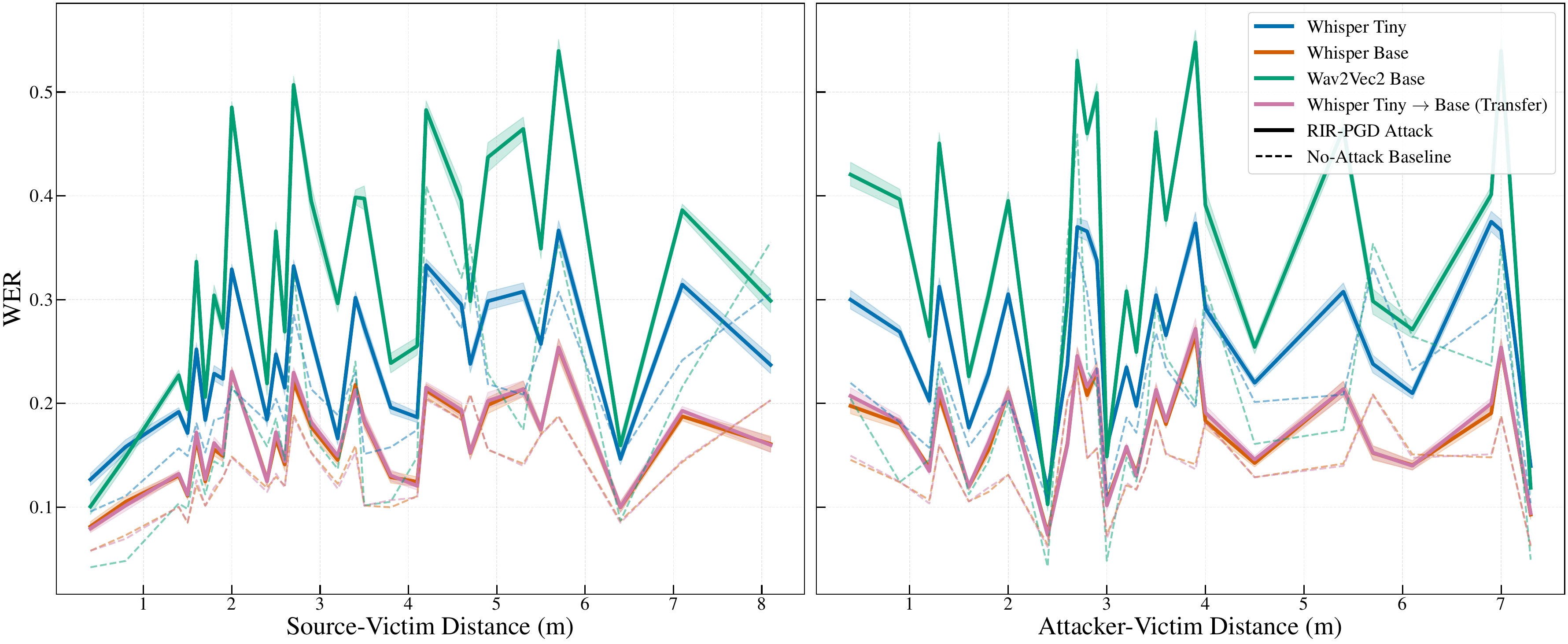}
      \caption{Impact of spatial geometry on WER. Success scales primarily with Source-Victim distance (acoustic distortion of the carrier) but remains largely invariant to
  Attacker-Victim distance, provided the attacker meets the energetic cost of projection.}
      \label{fig:dist_grid}
  \end{figure*}

While Figure~\ref{fig:dist_grid} exhibits a relatively chaotic dynamic between individual Source--Victim distances and the resulting WER, the cumulative distribution in Figure~\ref{fig:SV_Distance} reveals a macroscopic physical truth. By smoothing the stochastic interference patterns of specific wall reflections, we observe a clear, monotonic decay in model performance as the acoustic path length increases.

Crucially, the high degree of correlation between the attacked (solid) and clean (dashed) curves indicates that the room’s natural reverberation acts as a barrier to model performance that is missed evaluating within a strictly digital context. Channel distortion and reverberations provide a level of added difficulty that the attacker is not bypassing, but is amplifying to successfully attack the model. 

That the efficacy of the perturbation is fundamentally linked to the degradation of the carrier signal's quality (due to the influence of distance, and increased scope for reverberation based impacts) underscores the necessity of our room-aware modeling, not just for adversarial research, but for acoustic works more broadly. 

\begin{figure*}%
      \centering
      \includegraphics[width=0.85\textwidth]{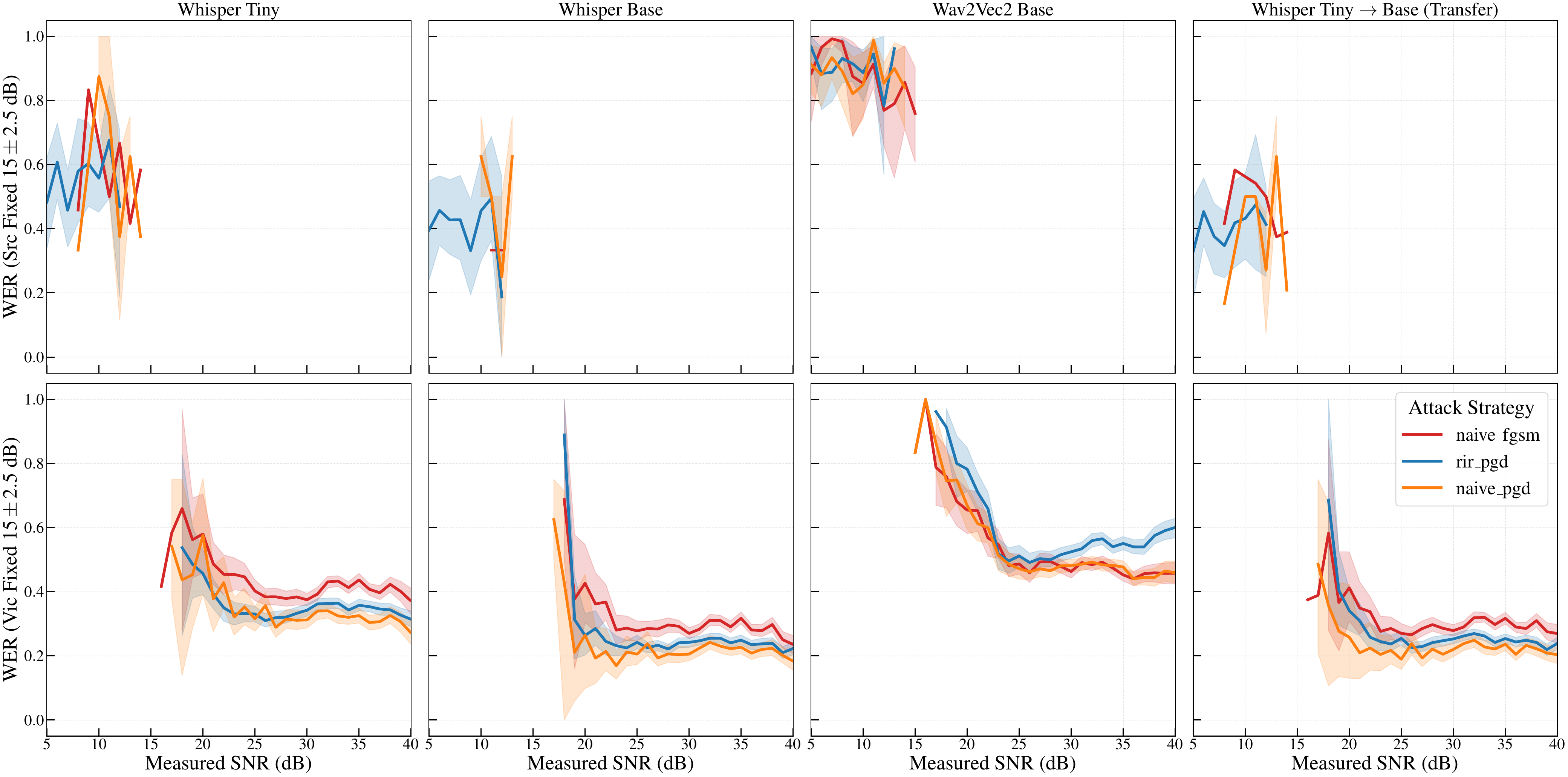}
      \caption{WER when the SNR is fixed at $15 \pm 2.5$ at either the source (top) or the victim (bottom).}
      \label{fig:WER_SRC_VIC_Fixed_15}
  \end{figure*}

\begin{figure*}[ht]
      \centering
      \includegraphics[width=0.85\textwidth]{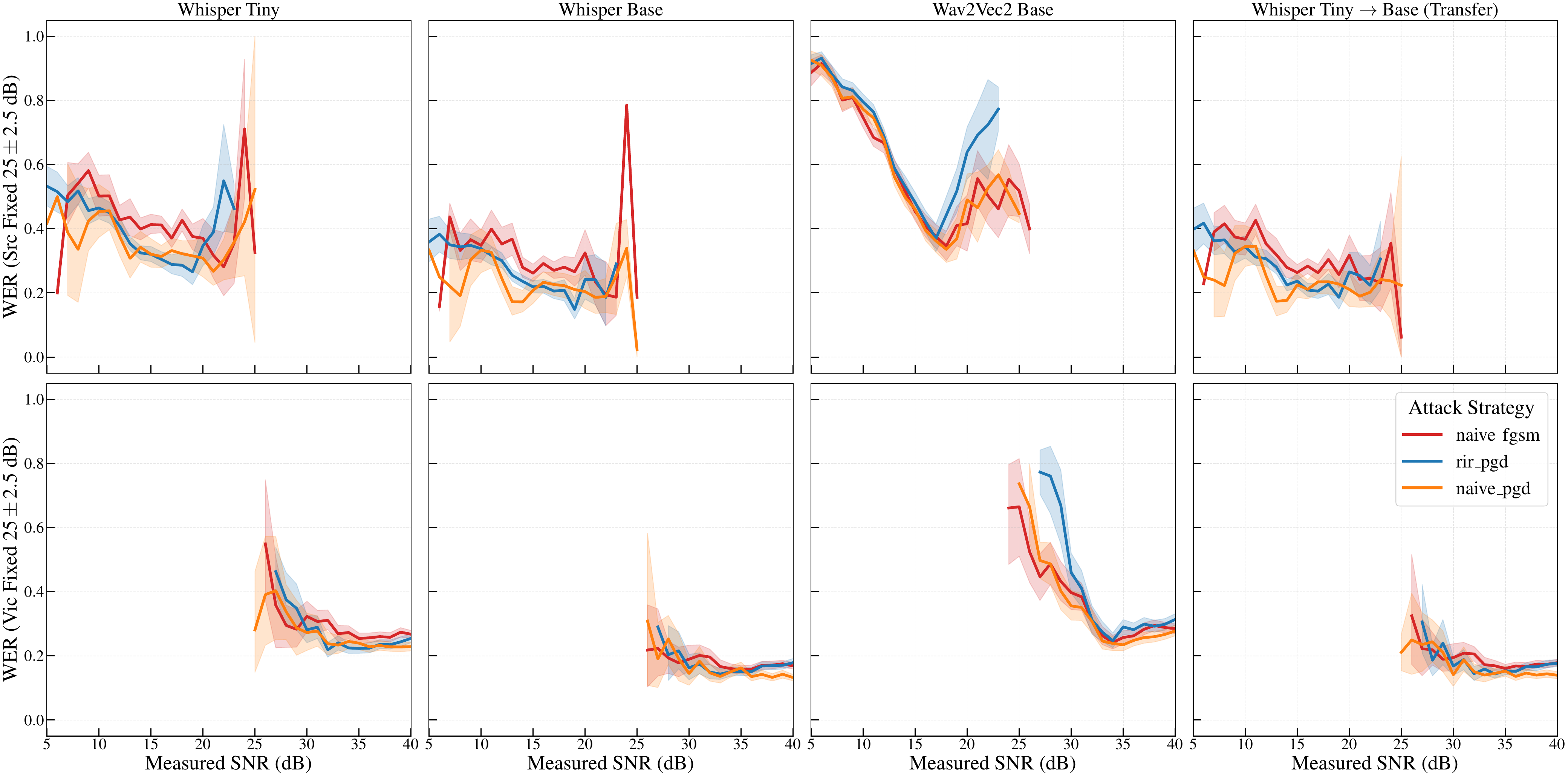}
      \caption{WER when the SNR is fixed at $25 \pm 2.5$ at either the source (top) or the victim (bottom).}
      \label{fig:WER_SRC_VIC_Fixed_25}
  \end{figure*}

Controlling for the SNR at source and victim in Figures~\ref{fig:WER_SRC_VIC_Fixed_15} and \ref{fig:WER_SRC_VIC_Fixed_25} presents a similar, but more complex picture, to that of Figure~\ref{fig:grad_grid}: the Naive FGSM typically outperforms all other approaches - producing higher WER's for the same perceived SNR. The exception to this is wav2vec, which appears to produce more consistent gradients, allowing the attack to navigate around the spectral nulls inherent to the acoustic environment.

\section{Extensions}\label{app:extensions}

We emphasize that our approach focuses upon establishing the metrological basis for rigorous, acoustically-aware explorations of adversarial risk in physically relevant environment. Prior to this, the field of audio AML was caught in a scale trap, where evaluations were either too fast to be physically relevant (in digital-only attacks), or too slow to be statistically rigorous (through physical OTA attack evaluations). By introducing a high-throughput, acoustically-aligned simulation framework, we provide the foundations required to consider more complex physical conditions. We now present options for future work, to expand upon these foundations.

\paragraph{Directivity} In an attempt to minimize the impact of observed variance, our simulations strictly considered speakers and microphones with linear responses. However, directional speakers and microphones have the potential to impact upon perceptions of the relative SNR, and the information available at each location. This could theoretically allow a sophisticated attacker to lower $\text{SNR}_s$ while maintaining a high $\text{SNR}_v$. 

While this could be considered to invalidate the Dual-Form SNR as a fixed environment property, the observation itself is only possible through our framework, which formally separates SNR at different locations. Furthermore, because our framework is modular, future researchers can consider directional radiation patterns directly into the RIR simulation framework we have provided, in order to provide an assessment of stealth under sophisticated attack conditions. 

\paragraph{Natural Perturbations} Similarly, the presence of non-stationary noise---of the kind seen through background chatter, traffic, or from other similar sources---may influence the Knowledge Gradient for the attacker. However, by providing an acoustically-aligned environment that is stable, repeatable, and scalable, this work enables the creation of such baseline scenarios, enabling researchers to quickly consider the layering of stochastic, real-world noise distributions on top of deterministic RIRs to study systemic robustness at scale, a task that would be physically impossible and digitally meaningless without the framework's architecture.

\paragraph{Scaling defenses} We acknowledge that the limited scope of the defenses (specifically 8-bit quantization), however we would stress that research into this space may well be the next frontier for the field. Similar to how this work demonstrates the importance of considering attacks in the context of the acoustic environment, it also opens up opportunities for new considerations of acoustically-aware defensive mechanisms.

\section{Computational Process}\label{ref:bottlenecks}

\begin{algorithm}[tb]
   \caption{Decoupled RIR-Aware Adversarial Attack. Prior to running this code, the set of $h$ is pre-computed on CPU and converted to GPU tensors. The $\text{Conv1D}$ operator allows for gradients $\nabla_{\delta}$ to flow through the network.}
   \label{alg:rir_attack}
\begin{algorithmic}
   \STATE {\bfseries Input:} Source audio $x$, Target text $T$, Model $f$, Knowledge Level $\mathcal{K}$
   \STATE {\bfseries Output:} Adversarial Example $x_{adv}$
   \STATE
   \STATE \COMMENT{Phase 1: Offline RIR Pre-computation (CPU)}
   \IF {$\mathcal{K} = \mathcal{K}_{oracle}$}
       \STATE $H_{batch} \leftarrow \{h_{true}\}$
   \ELSIF {$\mathcal{K} = \mathcal{K}_{blind}$}
       \STATE $H_{batch} \leftarrow \text{Sample} \ \{h_1, \dots, h_N\} \sim \mathcal{D}_{room}$
   \ENDIF
   \STATE Convert $H_{batch}$ to GPU Tensors
   \STATE Sample and convert $H_{batch}$, comprising the room RIR's $h$ to GPU Tensors
   \STATE
   \STATE \COMMENT{Phase 2: Online Optimization (GPU)}
   \STATE $x_t \leftarrow x$ \COMMENT{Initialize $x_0 = x$}
   \STATE
   \FOR{step $t=0$ {\bfseries to} $K-1$}
       \STATE \COMMENT{Calculate gradient to maximize loss relative to original text T}
       \STATE $g \leftarrow \nabla_{x_t} \mathcal{L}_{\text{ASR}}(f(x_t), T)$ 
       \STATE
       \STATE \COMMENT{Apply PGD update rule: $x_t + \alpha \cdot \text{sign}(g)$}
       \STATE $x_{t+1} \leftarrow x_t + \alpha \cdot \text{sign}(g)$
       \STATE
       \STATE \COMMENT{Projection operator $\Pi_{x+\mathcal{S}}$: Clip to $\epsilon$-ball and valid data range}
       \STATE $x_{t+1} \leftarrow \text{Clip}(x_{t+1}, x - \epsilon, x + \epsilon)$
       \STATE $x_{t+1} \leftarrow \text{Clip}(x_{t+1}, -1, 1)$ 
   \ENDFOR
   \STATE
   \STATE $x_{adv} \leftarrow x_K$
\end{algorithmic}
\end{algorithm}

An aim of this work is to provide a robust and generalizable framework for evaluating adversarial attacks against ASR systems, the approach for which is outlined within Algorithm~\ref{alg:rir_attack}. Within this work, we performed evaluations against the LibriSpeech \texttt{test-clean} English corpora~\citep{panayotov2015librispeech}, where all audio was normalized to 16 kHz mono signals, and filtered to ensure that no sample ran longer than $15$~seconds. We tested against Whisper (Tiny \& Base variants, MIT License, \citet{radford2023robust}) and Wav2Vec (Base variant, MIT License, \citet{baevski2020wav2vec}). Of these, Whisper is a sequence-to-sequence transformer model, while Wav2Vec Connectionist Temporal Classification (CTC)-based model. 

For our tested attacks, PGD was run over 40 iterations, with $\alpha=0.01$ and random starting, while FGSM was a single-step gradient sign update. The tested Carlini \& Wagner style attack employed $\ell_2$-loss optimization with 5 binary search steps for the trade-off constant $c$ and 100 iterations per step, with the learning rate set to $0.01$. 

As was discussed in the main body of the work, the tested room simulations are performed over thousands of randomized shoebox rooms whose lengths and widths are uniformly distributed over $[3,10]\text{m}$, and with a height between $[2.5, 3.5]\text{m}$. While our framework can be extended beyond shoebox room geometries, we deliberately restricted ourselves to such a modeling framework for two key criteria. The first relates to validation: RIR simulations are most strongly validated within the shoebox geometry context. The second is that non-uniform geometries present significant challenges with regards to identifying the characteristic lengths of acoustic signals, in order to appropriately contextualize results. With both of these points made, we stress that it is still possible for our framework to be extended to more complex geometries, however we leave solving these associated problems for future work.

\paragraph{Performance} The high-throughput performance of the system described within this work is a key part of being able to reach $8$~million adversarial attacks.  Table~\ref{tab:compute_breakdown} demonstrates that the workload is heavily skewed toward CPU-bound signal processing (PESQ/STOI) rather than GPU-bound adversarial optimization. Under optimal conditions, the total cost is estimated at $11.75$ GPU-Days. However, large-scale deployment on shared clusters resulted in an observed runtime of ${\sim}100$ GPU-Days. This $8.5\times$ discrepancy highlights a synchronization penalty where GPUs are starved while waiting for CPU-bound metrics.%

To again reiterate the implications of digital simulation, consider the time associated with just playing $1.28$ million samples directly within a single acoustic environment. Even if each sample was capped at $10$ seconds of length, this would require $925$ days of active time. That we are able to reach these scales within our acoustically-aligned digital simulation framework represents a significant step-change in the ability to approach physically realistic scenarios for evaluations of not just acoustic AML problems, but of the broader class of ASR inspired problems. 

\begin{table}[h]
\caption{Computational cost breakdown for the full evaluation suite. The processing of quality metrics constitutes the majority of the theoretical runtime. Costs incurred on a system with 64GB of ram and a 80GB A100 GPU.}
\label{tab:compute_breakdown}
\begin{center}
\begin{small}
\begin{sc}
\begin{tabular}{lcrr}
\toprule
Component & Count & Time & Load \\
\midrule
Attacks (Gen) & 1.28M & 2.45 Days & 20.9\% \\
Transcription & 10.2M & 1.35 Days & 11.5\% \\
Metrics (CPU) & 40.9M & 7.95 Days & 67.7\% \\
\midrule
\textbf{Total} & \textbf{-} & \textbf{11.75 Days} & \textbf{100\%} \\
\bottomrule
\end{tabular}
\end{sc}
\end{small}
\end{center}
\end{table}

\section{Defenses}\label{app:defenses}

A common defensive methodology is to consider the impact of a pre-model quantization phase. Quantization inherently reduces the information content of the signal, and thus has been posited to serve as a viable defense against adversarial manipulation. However, as can be seen in Table~\ref{tab:quantization}, there is a consistent increase (on the order of 10 $\%$) in the  observed WER. We posit that this likely stems from the decreased acoustic energy available at the victim receiver after factoring in $h_{SV}$, leading to the clean signal occupying only a fraction of the dynamic range. Thus quantization at this point has a disproportionate effect on the signal output, relative to past works that have strictly considered quantization within the digital domain. Counterintuitive to expectations, quantization acts as what is essentially a second broadband noise source, that increases the model's confusion. 

These results are reinforced by the differences in performance between wav2vec's CTC model, relative to Whispers Transformer architecture. We believe that quantization introduces variations that elicit a collapse in CTC alignment. By contrast the Transformer's attention mechanism allows for aggregation of information across longer context windows, increasing the models resilience to the impacts of quantization.

\begin{table*}[htbp]
\centering
\caption{Word Error Rates when considering both no defence, and 8-bit quantization as a defence. }\label{tab:quantization}
\begin{small}
\begin{sc}
\begin{tabular}{llccc}
\toprule
 & defense & No Defense & 8-bit Quant & Rel. Change ($\%$) \\
\midrule
\multirow[t]{3}{*}{Whisper Tiny} & Naive FGSM & 0.25 & 0.278 & +11.1$\%$ \\
 & Naive PGD & 0.221 & 0.257 & +16.1\% \\
 & RIR PGD & 0.25 & 0.28 & +11.9\% \\
\cline{1-5}
\multirow[t]{3}{*}{Whisper Base} & Naive FGSM & 0.168 & 0.184 & +9.6$\% $\\
 & Naive PGD & 0.144 & 0.166 & +15.5\% \\
 & RIR PGD & 0.168 & 0.185 & +10.5\% \\
\cline{1-5}
\multirow[t]{3}{*}{Wav2Vec2 Base} & Naive FGSM & 0.285 & 0.371 & +30.2$\%$ \\
 & Naive PGD & 0.274 & 0.377 & +37.3$\%$ \\
 & RIR PGD & 0.337 & 0.413 & +22.6$\%$ \\
\cline{1-5}
\multirow[t]{3}{*}{Whisper Tiny $\rightarrow$ Base (Transfer)} & Naive FGSM & 0.17 & 0.187 & +10.1$\%$ \\
 & Naive PGD & 0.146 & 0.169 & +15.8$\% $\\
 & RIR PGD & 0.17 & 0.188 & +10.5$\%$ \\
\cline{1-5}
\bottomrule
\end{tabular}
\end{sc}\end{small}
\end{table*}

\section{Acoustic Measures of Perceptual Quality}

\begin{figure*}%
    \centering
    \includegraphics[width=0.85\textwidth]{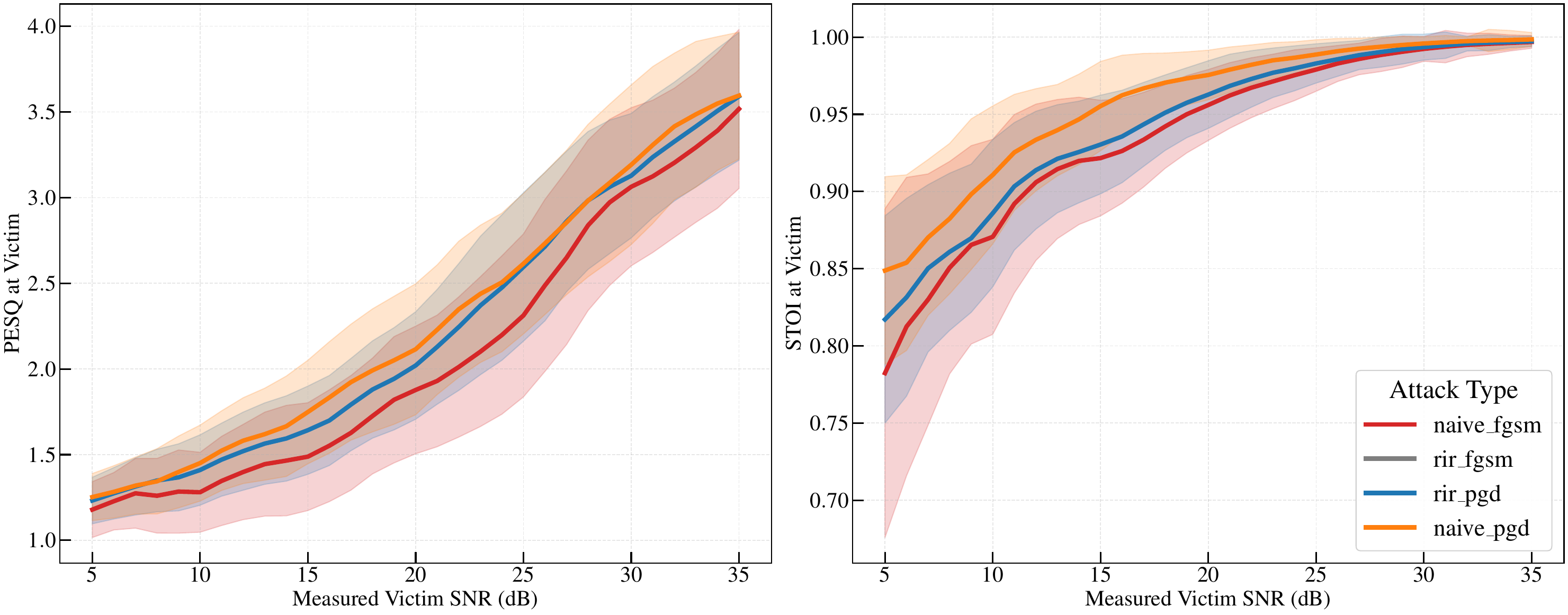}
    \caption{Perceptual measures of quality, broken down by attack type.}
    \label{fig:quality_attack}
\end{figure*}

It is important to note that while the performances of these attacks can be similar, there is a definite decoupling in terms of their perceptibility, with Figure~\ref{fig:quality_attack} demonstrating that Naive PGD consistently produces the largest PESQ and STOI scores~-- indicating a small degree of added stealth, relative to the other attacks. While the RIR PGD and Naive FGSM are almost equivalent for STOI, RIR PGD is relatively favoured for PESQ, and equals Naive PGD at higher Victim SNR levels. These results suggest that the RIR-aware attacks may appear to be slightly more natural than their FGSM equivalents, although these results must be tempered by our earlier discussions relating to the pitfalls of these perceptual measures. 

  \begin{figure}%
      \centering
      \includegraphics[width=0.325\textwidth]{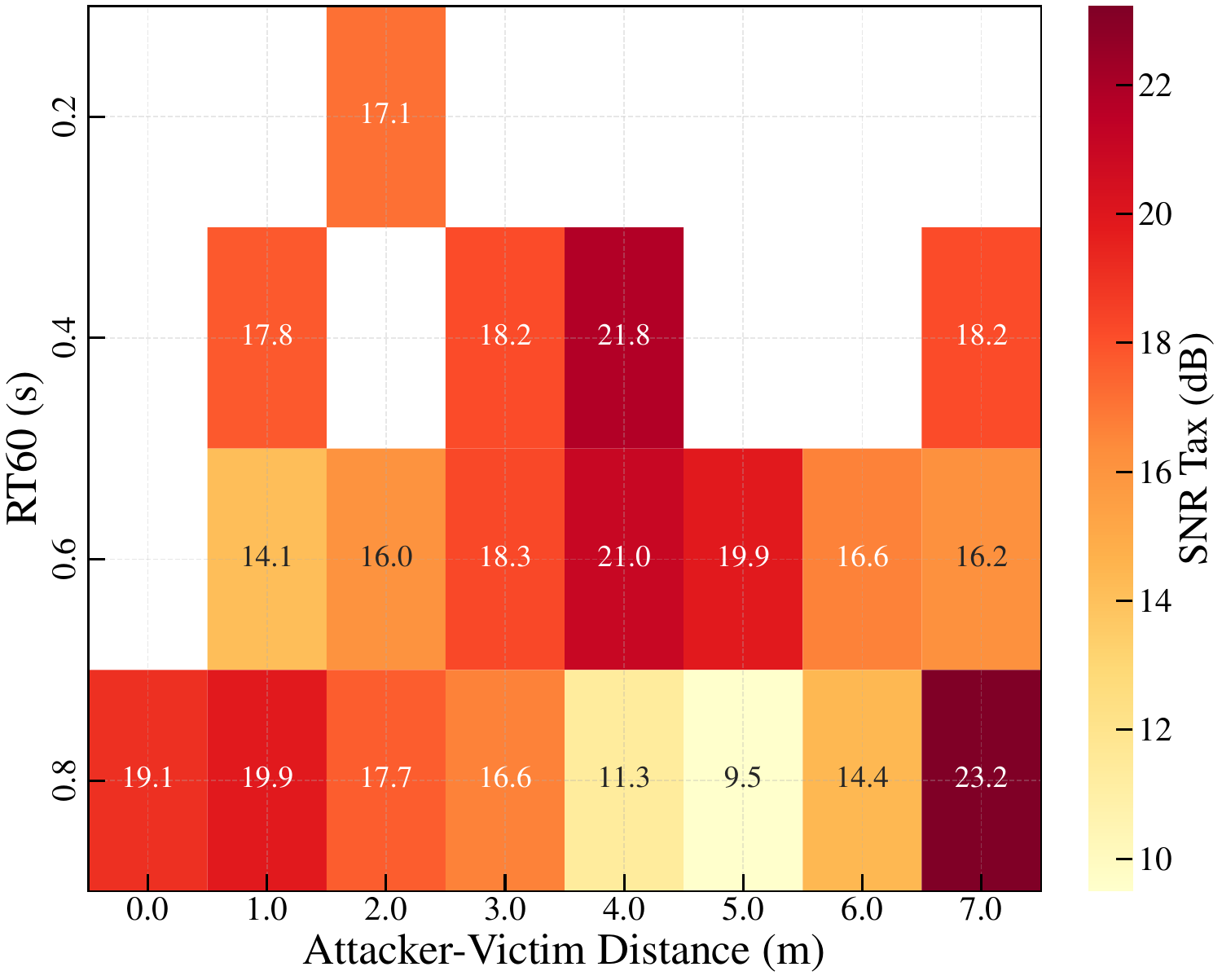}
      \caption{Acoustic Tax Heatmap, covering the SNR differential ($\text{SNR}_\text{source} - SNR_\text{victim}$) in terms of both $RT_{60}$ and Attacker-Victim distance, revealing how the energetic signature of an attack is non-uniformly distributed across the acoustic space.}
      \label{fig:tax_heatmap}
  \end{figure}

  \begin{figure*}%
      \centering
      \includegraphics[width=0.85\textwidth]{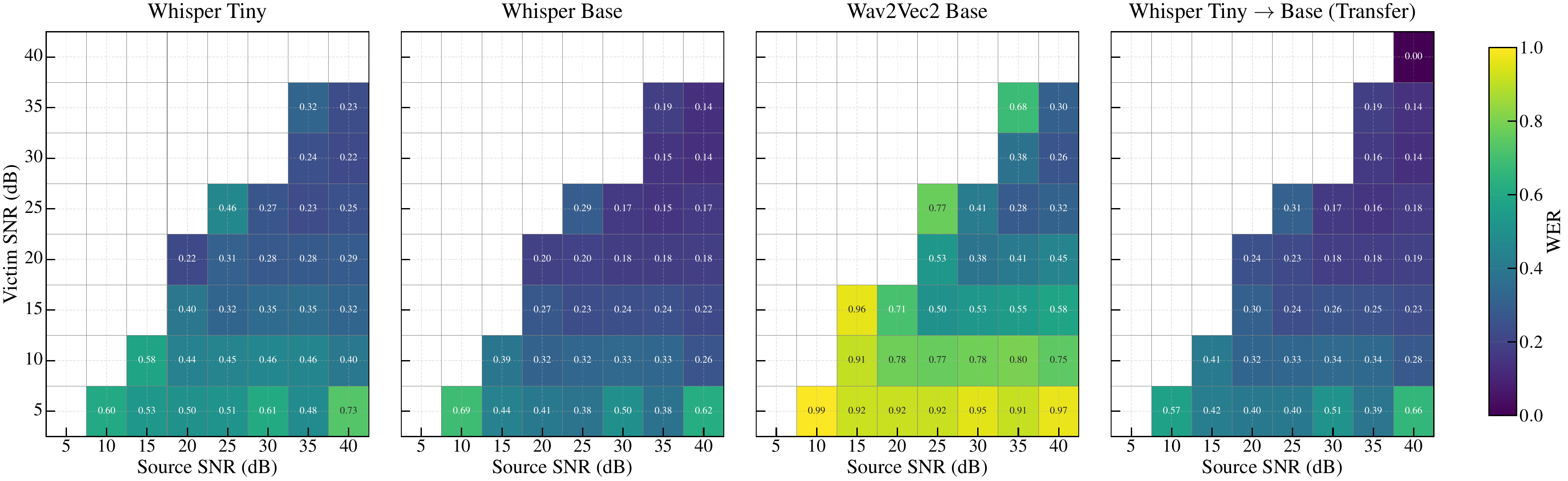}
      \caption{Relationship between the dual-SNR metrics and WER.}
      \label{fig:geometry_of_WER}
  \end{figure*}

\section{Defining $c$}\label{app:process}

Our aim with $c$ in Equation~\ref{eqn:constraint} was not to define a hard constraint, that exactly mapped to a particlar source SNR. Rather, our approach was to seed our experimental set with a range of source SNRs, that roughly existed within a range of values. To achieve this, we began by sampling a set of $10$ random acoustic environments. Each of these rooms were exposed to $100$ samples at varying $c$, creating a mapping between the source SNR and $c$. We then selected a range of $c$ values associated with SNRs of $\{15, 25, 35, 45\}$, and used these values to guide our attacks. 

We again emphasize that the exact relationship between $c$ and the SNR is not necessary for our analysis, and would, in fact, produce unfavorable outcomes. Our approach involves exploring a large range of acoustic environments, each of which would have their own unique mapping between these two parameters. Thus a single value of $c$ produces a range of SNR values at the source, and, commensurately, at the victim as well. Thus our choice to use a set of $c$ values was a deliberate one, to ensure that our environments demonstrated a sufficient diversity of SNR values.

\end{document}